\documentclass[preprint,12pt,3p]{elsarticle}

\usepackage{graphics}
\usepackage{graphicx}
\usepackage{epsfig}

\usepackage{amssymb}
\usepackage{xcolor}
\usepackage{times}
\usepackage{epsfig}
\usepackage{graphicx}
\usepackage{subfig}
\usepackage{float}
\usepackage{amsmath}
\usepackage{amssymb}
\usepackage{algorithm}
\usepackage{algorithmic}
\usepackage{multirow}
\usepackage{setspace}
\makeatletter \@fpsep = 2pt \makeatother
\begin{document}

\begin{spacing}{2.0}

\begin{frontmatter}

\title{Multi-stage image denoising with the wavelet transform}

\author[label1,label2]{Chunwei Tian\corref{cor1}}
\ead{chunweitian@nwpu.edu.cn}
\author[label1]{Menghua Zheng}
\author[label3,label4]{Wangmeng Zuo}
\author[label5]{Bob Zhang}
\author[label6,label2]{Yanning Zhang\corref{cor1}}
\ead{ynzhang@nwpu.edu.cn}
\cortext[cor1]{Corresponding author}
\author[label7,label8]{David Zhang}

\address[label1]{School of Software, Northwestern Polytechnical University, Xi’an, Shaanxi, 710129, China }
\address[label2]{National Engineering Laboratory for Integrated Aero-Space-Ground-Ocean Big Data Application Technology, Xi’an, Shaanxi, 710129, China}
\address[label3]{School of Computer Science and Technology, Harbin Institute of Technology, Harbin, Heilongjiang, 150001}
\address[label4]{Peng Cheng Laboratory, Shenzhen, Guangdong, 518055, China}
\address[label5]{Department of Computer and Information Science, University of Macau, Macau, 999078, China}
\address[label6]{School of Computer Science, Northwestern Polytechnical University, Xi’an, Shaanxi, 710129, China}
\address[label7]{School of Data Science, The Chinese University of Hong Kong (Shenzhen), Shenzhen, 518172, Guangdong, China}
\address[label8]{Shenzhen Institute of Artificial Intelligence and Robotics for Society, Shenzhen, China}

\begin{abstract}
Deep convolutional neural networks (CNNs) are used for image denoising via automatically mining accurate structure information. However, most of existing CNNs depend on enlarging depth of designed networks to obtain better denoising performance, which may cause training difficulty. In this paper, we propose a multi-stage image denoising CNN with the wavelet transform  (MWDCNN) via three stages, i.e., a dynamic convolutional block (DCB), two cascaded wavelet transform and enhancement blocks (WEBs) and a residual block (RB). DCB uses a dynamic convolution to dynamically adjust parameters of several convolutions for making a tradeoff between denoising performance and computational costs. WEB uses a combination of signal processing technique (i.e., wavelet transformation) and discriminative learning to suppress noise for recovering more detailed information in image denoising. To further remove redundant features, RB is used to refine obtained features for improving denoising effects and reconstruct clean images via improved residual dense architectures. Experimental results show that the proposed MWDCNN outperforms some popular denoising methods in terms of quantitative and qualitative analysis. Codes are available at https://github.com/hellloxiaotian/MWDCNN.
\end{abstract}


\end{frontmatter}

% \linenumbers

\section{Introduction}
\label{sec-1}

Noisy images often arise in the high-level vision tasks, which makes image denoising become an important task in the field of low-level vision \cite{tian2020deep}. Specifically, a clean image as well as ${x}$ of image denoising problem can be represented via a degradation model of ${x=y-n}$, where ${y}$ and ${n}$ stand for a given noisy image and additive white Gaussian noise (AWGN) of standard deviation ${\sigma}$, respectively \cite{tian2020deep}.  According to that, scholars developed a lot of denoisers \cite{wang1999progressive}. For instance, Wang et al. combined a switching scheme and progressive methods based median filter to obtain more detailed information for salt-pepper image denoising \cite{wang1999progressive}.  
To improve denoising performance, prior knowledge was used to extract useful information in image denoising \cite{rabbani2009image}. Rabbani et al. depended on Laplacian random variables of high local correlation to obtain an estimator based maximum a posteriori and minimum mean squared for image denoising \cite{rabbani2009image}. To improve denoising effect, mapping similar 2D-fragments of corrupted images into 3D-data arrays  can enhance sparsity in image denoising \cite{dabov2007image}. Besides, using different gradient vectors to conduct  absolute cosine value can recover detailed information \cite{li2022joint}. Also, an edge-preserving regularization term is used to overcome adverse effect of unreliable prior from  a guidance image for image denoising \cite{li2022joint}. Although mentioned these methods have performed well in image denoising, they suffered from two shortages as follows \cite{zhang2017beyond}. That is, they resorted to manually selecting parameters. Also, they may require complex optimization algorithms to pursue excellent denoising effects. 

To address the problem, discriminative learning methods in end-to-end ways rather than manually tuning parameters are developed \cite{tian2020deep}. Due to powerful learning abilities, discriminative learning methods, especially convolutional neural networks (CNNs) are applied to resolve image denoising problem \cite{zhang2017beyond}.  Zhang et al. exploited combinations of stacked convolutions, batch normalization (BN) and ReLU to implement an efficient CNN denoiser \cite{zhang2017beyond}. To obtain salient features, attention idea is embedded into a network in image processing application \cite{tian2020attention}. For instance, Tian et al. used an attention mechanism to guide a CNN to extract more accurate information for image denoising \cite{tian2020attention}.  To improve denoising performance, making full use of hierarchical information can facilitate more useful information to suppress noise \cite{tai2017memnet}.  For instance, Tai et al. utilized recursive units and gate units to transfer memory abilities of shallow layers to deep layers to improve quality of restored images \cite{tai2017memnet}. To further improve performance in image restoration, Zhang et al. used residual dense blocks in a deeper CNN to integrate local and global features for obtaining more robust features in image restoration \cite{zhang2018residual}. Although these methods are effective in image denoising, they may rely on deeper architectures to pursue excellent denoising performance, which may increase difficulty of training. In this paper, we present a multi-stage image denoising CNN with the wavelet transform as well as MWDCNN. It relies on three stages, i.e., a dynamic convolutional block (DCB), two cascaded stacked wavelet transform and enhancement blocks (WEBs) and a residual block (RB). DCB dynamically adjusts parameters of several convolutions via a dynamic convolution to overcome drawbacks of poor performance of some lightweight CNNs in terms of network depth and width for making a tradeoff between denoising performance and computational costs. Stacked WEBs use combinations of signal processing techniques (i.e., wavelet transform) and discriminative learning (i.e., residual dense blocks) to suppress noise for recovering more detailed information in image denoising. To further remove redundant features, RB is used to refine obtained features for improving denoising effects and reconstruct clean images via improved stacked residual dense architectures. Besides, the proposed MWDCNN is superior to popular denoising methods, i.e., a denoising CNN (DnCNN) \cite{zhang2017beyond} and attention-guided denoising (ADNet) \cite{tian2020attention} in terms of quantitative and qualitative analysis. 

Contributions of this paper can be shown as follows. 

(1)	 A dynamic convolution is used into a CNN to address limitations in depth and 
width of lightweight CNNs for pursuing good denoising performance.   

(2)	The combination of a signal processing technique and discriminative learning 
technique is used for image denoising. 

(3)	Enhanced residual dense architectures are used to remove redundant 
information for improving denoising effects. 

The remainder of this paper is organized as follows. Section 2 provides related work about deep CNNs for image denoising, dynamic convolutions and wavelet transform techniques. Section 3 describes the proposed denoising method. Section 4 shows datasets, experimental settings, the proposed method analysis and extensive experimental results. Section 5 reports conclusion of this paper. 

\section{Related work}
\subsection{Deep CNNs for image denoising}
To overcome drawbacks of traditional machine learning in image denoising, networks based components, i.e., CNNs are proposed \cite{zhang2018ffdnet}. To accelerate training speed of denoisers, Zhang et al. used patches of noisy images and noisy mapping to act a CNN to achieve an efficient denoising network \cite{zhang2018ffdnet}. To improve denoising performance, combining dilated convolutions and BN in a CNN can obtain more accurate context information to promote denoising effect \cite{tian2020deep}.  Mentioned methods have obtained excellent denoising performance, however, convolutional kernels of these methods have same weights for noisy images, which may have big computational costs. Differing from these methods, our denoising method uses dynamic convolutions in a CNN to adjust parameters of convolutional kernels to train a robust denoiser, where mentioned dynamic convolutions can be shown in Section 2.2. Besides, we combine frequency features via wavelet transform and structural information via a CNN to obtain complementary information in image denoising, where mentioned wavelet transform can be shown in Section 2.3. 

\subsection{Dynamic convolutions for image applications}
Most of existing methods share parameters of each convolutional layer to train a denoising CNN for image applications (i.e., image classification), however, they cannot adjust parameters of each layer to obtain a robust classifier, according to different images \cite{chen2020dynamic}. To address this question, dynamic convolutions used multiple parallel convolutions via attention mechanisms are fused to reduce computational costs of different CNNs for training a robust classifier \cite{chen2020dynamic}. To improve training efficiency, a dynamic convolution was embedded into a CNN to remove redundancy features for improving classification accuracy \cite{zhang2020dynet}. To mine more context, Sun et al. used Gaussian dynamic pyramid pooling in a CNN to improve expressive abilities in image segmentation \cite{sun2021gaussian}.  Besides, using matrix decomposition to guide a dynamic convolution can make a tradeoff between performance and  training speed in image recognition \cite{li2021revisiting}. Due to excellent performance, a dynamic convolution is used into a CNN for image denoising in this paper. 

\subsection{Wavelet transform for image applications}
It is known that images can be treated as signals, thus, signal processing techniques, i.e., wavelet transform are effective for low-level tasks \cite{cho2005multivariate}. For instance, Cho et al. used multivariate statistical idea to estimate coefficients of wavelet transform for filtering noise in image denoising \cite{cho2005multivariate}. Liu et al. utilized wavelet transform and genetic algorithm to suppress noise in image denoising \cite{liu2015image}. To mine more useful information, wavelet transform is used into a CNN to learn detailed information and content information for image super-resolution \cite{guo2017deep}. As an alternative, using cross-connection and residual technique to integrate a CNN and wavelet transform was a good tool for image super-resolution \cite{yang2021effective}. According to mentioned illustrations, we can see that wavelet transform is effective for low-level task. Inspired by that, we fuse two cascaded wavelet transform techniques into a CNN to implement fusion of frequency information and struct information for promoting visual denoising effects. 
%%%%%%%%%%%%%%%%%%%%%%%%%%%%%%%%%%%%%%%%%%%%%%%%%%%%%%%%%%%%%%%%%%%%%%%%%%%%%%%%%%%%%%%%%%
%%%%%%%%%%%%%%%%%%%%%%%%%%%%%%%%%%%%%%%%%%%Fig1%%%%%%%%%%%%%%%%%%%%%%%%%%%%%%%%%%%%%%%%%%%
%%%%%%%%%%%%%%%%%%%%%%%%%%%%%%%%%%%%%%%%%%%%%%%%%%%%%%%%%%%%%%%%%%%%%%%%%%%%%%%%%%%%%%%%%%
\begin{figure}
\centering
\includegraphics[width=6in, keepaspectratio]{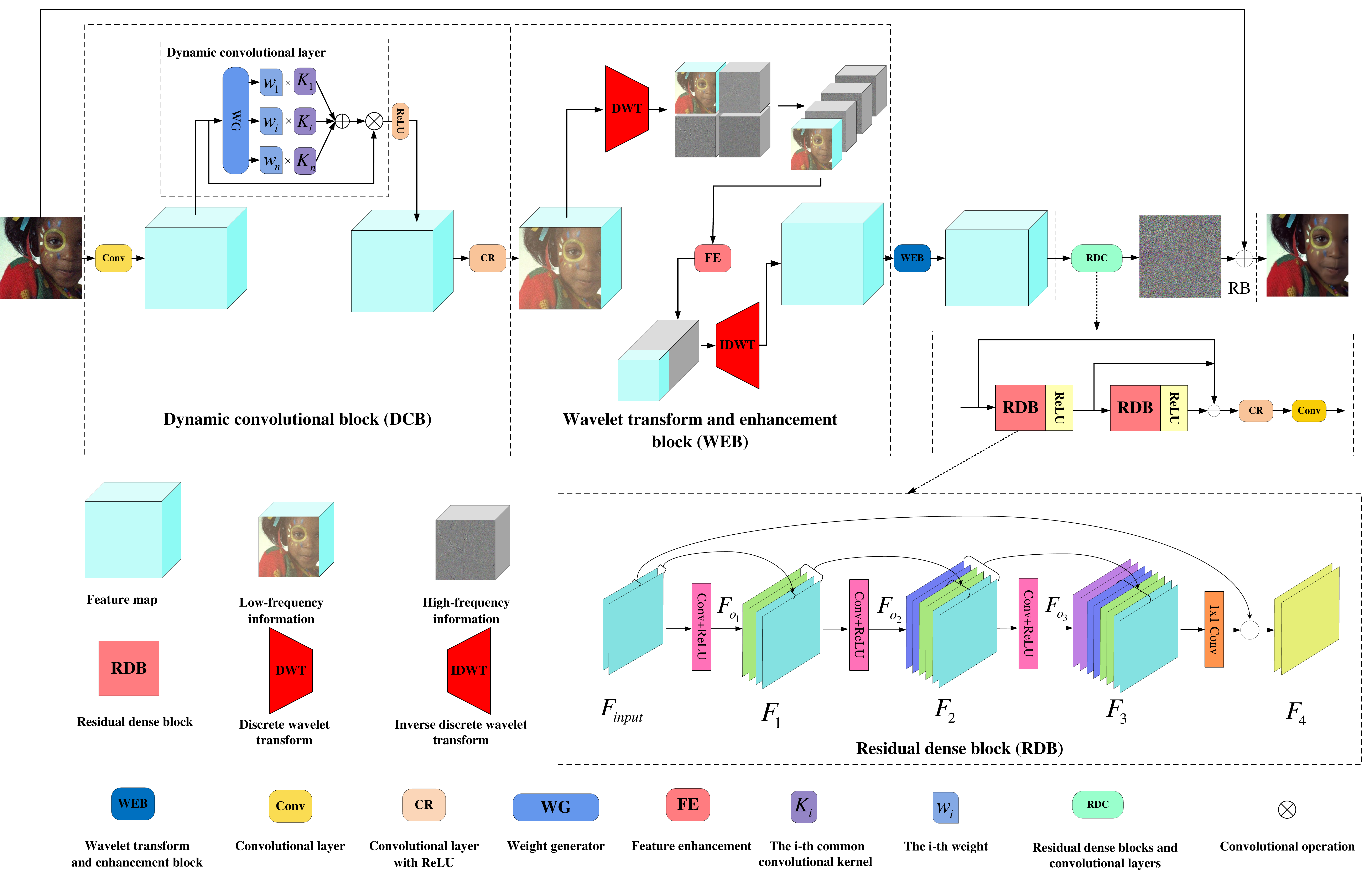}\\
\caption{Network architecture of MWDCNN.}
\end{figure}

\section{The proposed method}
\subsection{Network architecture}

The designed 23-layer MWDCNN contains three parts, i.e., a DCB, two cascaded WEBs and a RB for image denoising as shown in Fig.1. A 5-layer DCB uses a dynamic convolution to dynamically adjust parameters of several convolutions to make a tradeoff between denoising performance and computational costs, according to different images. To extract more useful information, two stacked 8-layer WEBs combine signal processing technique and discriminative learning to remove noise. That is, it combines frequency features via wavelet transform and structure information via convolutional layers to mine robust features in image denoising. Also, a 10-layer RB is used to refine obtained features via enhanced residual dense architectures and reconstruct clean images via a residual learning operation. To vividly express the mentioned process, the following equation is conducted. 

\begin{small}
\begin{equation}
\begin{array}{ll}
{I_{C}} & =  {f_{RB}(f_{WEB}(f_{WEB}(f_{DCB}(I_N))))}\\
 & =   {f_{MWDCNN}(I_{N})},
\end{array}
\end{equation}
\end{small}
 where ${I_N}$ and ${I_C}$ denote given a noisy image and a clean image, respectively. ${f_{DCB}}$ stands for a function of DCB. ${f_{WEB}}$ and ${f_{RB}}$ are functions of WEB and RB, respectively. ${f_{MWDCNN}}$ expresses a function of MWDCNN. MWDCNN is trained by a loss function in Section 3.2.

\subsection{Loss function}
To keep consistency with popular denoising methods, i.e., DnCNN \cite{zhang2017beyond} and fast and flexible denoising network (FFDNet) \cite{zhang2018ffdnet}, we choose a mean square error (MSE) \cite{allen1971mean} as a loss function (also treated as objective function) to train a denoiser of MWDCNN. Specifically, MSE uses pairs of ${ \{I_C^i,I_N^i\}}$ (${1=<i<=n}$) in a supervised way to train this denoising model, where $I_C^i$ and $I_N^i$ are defined as the ${i-th}$ clean image and noisy image, respectively.  ${n}$ represents the number of noisy images in the training process. MWDCNN can be optimized by Adam to obtain suitable parameters \cite{kingma2014adam}. Besides, to show its good performance of MSE, we discuss validity of different losses from Charbonnier of the newest denoising method\cite{liang2021swinir} and Pearson of quality assessment method  \cite{ayyoubzadeh2021asna} fused into our MWDCNN for image denoising in this Section 4.3. The mentioned procedure can be formulated as Eq. (2). 

\begin{small}
\begin{equation}
\begin{array}{ll}
l\left( \theta  \right){\rm{ }} = {\rm{ }}\frac{1}{{2n}}\sum\limits_{i = 1}^n {||{f_{MWDCNN}}(I_N^i) - (I_C^i)|{|^2}} ,
\end{array}
\end{equation}
\end{small}
where ${l}$ and ${\theta }$  express loss function and parameter sets.

\subsection{Dynamic convolution block}
The first stage of MWDCNN is implemented by a dynamic convolution block. The dynamic convolution block is composed of five layers: a convolutional layer, a 3-layer dynamic convolution and a convolutional layer. The first convolutional layer is used to convert noisy images to linear features, where its parameters are input channel number of 3 or 1 (depending on whether the input is a color or gray noisy images), convolutional kernel of  ${5\times 5}$ and output channel number of 64. The 3-layer dynamic convolution consists of 2-layer weight generator and 1-layer convolutional layer with ${5 \times 5}$ , where the weight generator contains an average pooling, a combination of convolutional layer with  ${1 \times 1}$ and ReLU, a convolutional layer with ${1 \times 1}$, and Softmax as illustrated in Fig. 2 \cite{chen2020dynamic}. The second convolutional layer followed by ReLU is used to refine obtained features from the dynamic convolution, where its parameters are input channel number of 64, convolutional kernel of ${5 \times 5}$  and output channel number of 64. These illustrations can be presented as follows.

\begin{small}
\begin{equation}
\begin{array}{ll}
O_{DCB} & = {f_{DCB}(I_N)}\\
& = CR(R(DC(C(I_N)))),
\end{array}
\end{equation}
\end{small}
where ${DC}$ denotes a function of dynamic convolution and ${O_{DCB}}$ is output of DCB, which acts two stacked WEBs. ${C}$, ${R}$ and  ${CR}$   stand for a convolution operation of  ${5 \times 5}$, activation function as well as ReLU and convolution operation of ${5 \times 5}$  with ReLU, respectively.

Implementations of mentioned dynamic convolution are shown as follows \cite{chen2020dynamic}. Firstly, it uses WG to obtain four weights to act four parallel convolutional kernels in a weighted way to adjust parameters. Secondly, obtained results act a convolutional layer of ${5 \times 5}$ . Thirdly, output of the first convolution in a DCB and output of second step in a dynamic convolution linearly fuse obtained features by obtaining suitable parameters for different noisy images. Work of a dynamic convolution in a DCB can be symbolled by the following Equations.

\begin{small}
\begin{equation}
\begin{array}{ll}

{O_{DC}} & = DC({O_{DCB\_1C}})\\
& =  C{\rm{(}}WG{\rm{(}}{O_{DCB\_1C}}{\rm{) }} \times K) \times {O_{DCB\_1C}}\\
&    = C{\rm{(}}\sum\limits_{i = 1}^4 {{\rm{W}}{{\rm{G}}_i}{\rm{(}}{O_{DCB\_1C}}{\rm{)}}} {\rm{ }} \times {K_i}) \times {O_{DCB\_1C}},

\end{array}
\end{equation}
\end{small}
where ${O_{DCB\_1C}}$ is an output of the first convolutional layer in the DCB. ${DC}$ denotes function of a dynamic convolution. ${WG}$, ${C}$ and  ${K}$ stand for functions of WG, a convolution of ${5 \times 5}$ and four parallel convolutional kernels, respectively. Also, this convolutional layer has input and output channel number of 64.  ${WG_i}$ and ${K_i}$ are used to represent the ${i-th}$ channel of results from WG and the ${i-th}$ convolutional kernel in four parallel convolutional kernels. ${O_{DC}}$ expresses an output of a dynamic convolution. Specifically, a function of WG \cite{chen2020dynamic} can be defined as Eq. (5).

\begin{small}
\begin{equation}
\begin{array}{ll}

{O_{WG}} &= WG{\rm{(}}{O_{DCB\_1C}}{\rm{) }}\\
\rm        &= S{\rm{(}}{C_1}(R{C_1}{\rm{(}}AP{\rm{(}}{O_{DCB\_1C}})))),

\end{array}
\end{equation}
\end{small}
where ${AP}$ and ${S}$ are defined as an average pooling and Softmax, respectively. $RC_1$ and $C_1$  are the combination of a convolutional layer of  ${1 \times 1}$ and a ReLU, and a convolutional layer of ${1 \times 1}$. The input channel number and output channel number of the first convolutional layer in the WG are 64 and 4, respectively. The input channel number and output channel number of the second convolutional layer in the WG are 4 and 4, respectively. Besides, WG can be visualized as Fig. 2. 
%%%%%%%%%%%%%%%%%%%%%%%%%%%%%%%%%%%%%%%%%%%%%%%%%%%%%%%%%%%%%%%%%%%%%%%%%%%%%%%%%%%%%%%%%%
%%%%%%%%%%%%%%%%%%%%%%%%%%%%%%%%%%%%%%%%%%%Fig2%%%%%%%%%%%%%%%%%%%%%%%%%%%%%%%%%%%%%%%%%%%
%%%%%%%%%%%%%%%%%%%%%%%%%%%%%%%%%%%%%%%%%%%%%%%%%%%%%%%%%%%%%%%%%%%%%%%%%%%%%%%%%%%%%%%%%%
\begin{figure}
\centering
\includegraphics[width=5in, keepaspectratio]{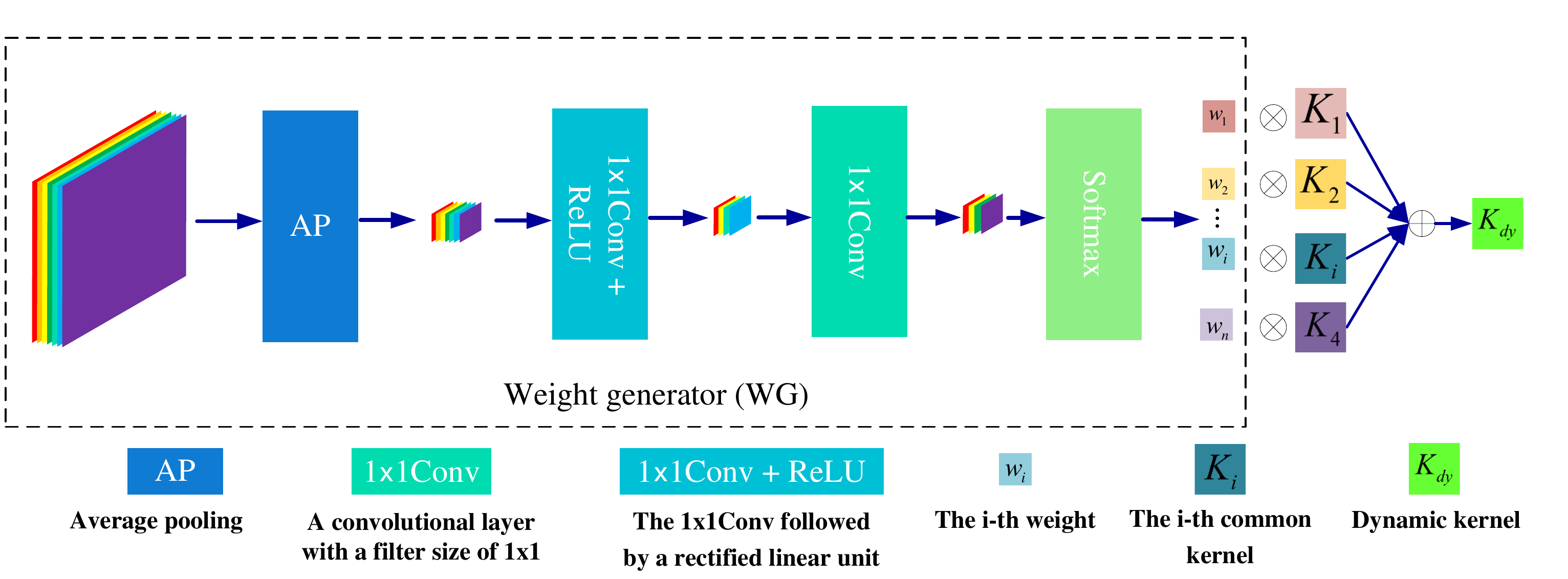}\\
\caption{Architecture of weight generator (WG).}
\end{figure}

\subsection{Two stacked wavelet transform and enhancement blocks}
The second phase of WDMCNN uses two stacked wavelet transform and enhancement blocks. Specifically, each 4-layer wavelet transform and enhancement block as well as WEB includes three steps to fuse frequency features and structure information via combining signal processing techniques and discriminative learning for obtaining more useful information. The first step uses discrete wavelet transform (DWT) \cite{feng2021optical} to convert linear construct information to four frequency features. The second step utilizes structural network to guide signal processing techniques via a feature enhancement (FE) mechanism (also treated as a 4-layer residual dense block (RDB) in Fig. 3 \cite{zhang2018residual}) to suppress noise for recovering more detailed information in image denoising. It is noted that sizes of all convolutional kernels besides the last covolutional kernel in a FE are ${5 \times 5}$ and the last convolutional kernel in a FE is $1\times1$. The third step uses an inverse discrete wavelet transform (IDWT) to convert frequency features to linear structure information. The mentioned illustrations can be shown in Eq. (6)

\begin{small}
\begin{equation}
\begin{array}{l}

%{O_{WEB}} &= {f_{WEB}}({f_{WEB}}({O_{DCB}}))\\
%{\rm{       }} &= {f_{WEB}}({f_{IDWT}}({f_{FE}}({f_{DWT}}({O_{DCB}}))))\\
%{\rm{       }} &= {f_{IDWT}}({f_{FE}}({f_{DWT}}({f_{IDWT}}({f_{FE}}({f_{DWT}}({O_{DCB}}))))))

\begin{array}{l}
\begin{split}
{O_{WEB}} &= {f_{WEB}}({f_{WEB}}({O_{DCB}}))\\
{\rm{       }} &= {f_{WEB}}({f_{IDWT}}({f_{FE}}({f_{DWT}}({O_{DCB}}))))\\
{\rm{       }} &= {f_{IDWT}}({f_{FE}}({f_{DWT}}({f_{IDWT}}({f_{FE}}({f_{DWT}}({O_{DCB}})))))),
\end{split}
\end{array}
\end{array}
\end{equation}
\end{small}
where ${f_{WEB}}$ expresses a function of WEB. ${f_{DWT}}$ and ${f_{IDWT}}$ denote functions of DWT and IDWT, respectively. ${f_{FE}}$ and ${O_{WEB}}$ represent function of FE and output of two stacked WEBs, respectively. More information of mentioned DWT and IDWT can be obtained in Ref. \cite{feng2021optical}. Also, mentioned 4-layer FE as well as a RDB contains three combinations of a convolutional layer and ReLU (also treaded as Conv+ReLU) and a convolutional layer (also treaded as Conv with ${1 \times 1}$) and its more detailed information can be obtained in Ref. \cite{zhang2018residual}.

\subsection{Residual block}
The RB is implemented a 10-layer enhanced residual dense architecture. That is, it    is composed of two combinations of residual dense block and ReLU and two convolutional layers as well as RDC, and two residual learning operations. That can be implemented by four steps as follows. The first step uses two combinations of a 4-layer RDBs \cite{zhang2018residual} and ReLU as the first component to further refine features in image denoising. To prevent long-term dependency problem, the second step uses two residual learning operations to fuse obtained features from two combinations of each RDB and ReLU, and WEB to enhance memory abilities of shallow layers on deep layers in image denoising. To prevent over-enhancement phenomenon of the second step, the third step uses a convolutional layer followed by ReLU to refine these features, where its input channel and output channel are 64, respectively. Finally, a convolutional layer is used to reconstruct noisy mappings and a residual learning operation is used to reconstruct clean images via obtained noisy mappings and given noisy images. izes of all convolutional kernels besides the fouth and eighth convolutions in a residual block are ${5\times5}$. Also, the fourth and eighth convolutional kernels are $1\times 1$. Input channel number and output channel number of two RDBs are 64, and input channel number of the last convolutional layer is 64. And the output channel number of the last convolutional layer is 3 or 1, depending on whether the noisy image is a color or gray images. The illustrations above can be shown in Eq. (7).

\begin{small}
\begin{equation}
\begin{array}{ll}
{I_C} &= {f_{RB}}({O_{WEB}})\\
&= {I_N} - {f_{RDC}}({O_{WEB}})\\
&= {I_N} - C(CR({O_{WEB}} + R({f_{RDB}}({O_{WEB}}))+R({f_{RDB}}(R({f_{RDB}}({O_{WEB}})))))),
\end{array}
\end{equation}
\end{small}
where ${f_{RDC}}$ denotes a function of RDC. Also, ‘-’ stands for a residual learning operation as well as ${\oplus }$ in Fig. 1.  ${f_{RDB}}$ and ${R}$ express a function of RDB and ReLU, respectively. 
%%%%%%%%%%%%%%%%%%%%%%%%%%%%%%%%%%%%%%%%%%%%%%%%%%%%%%%%%%%%%%%%%%%%%%%%%%%%%%%%%%%%%%%%%%
%%%%%%%%%%%%%%%%%%%%%%%%%%%%%%%%%%%%%%%%%%%Fig3%%%%%%%%%%%%%%%%%%%%%%%%%%%%%%%%%%%%%%%%%%%
%%%%%%%%%%%%%%%%%%%%%%%%%%%%%%%%%%%%%%%%%%%%%%%%%%%%%%%%%%%%%%%%%%%%%%%%%%%%%%%%%%%%%%%%%%
\begin{figure}
\centering
\includegraphics[width=6in, keepaspectratio]{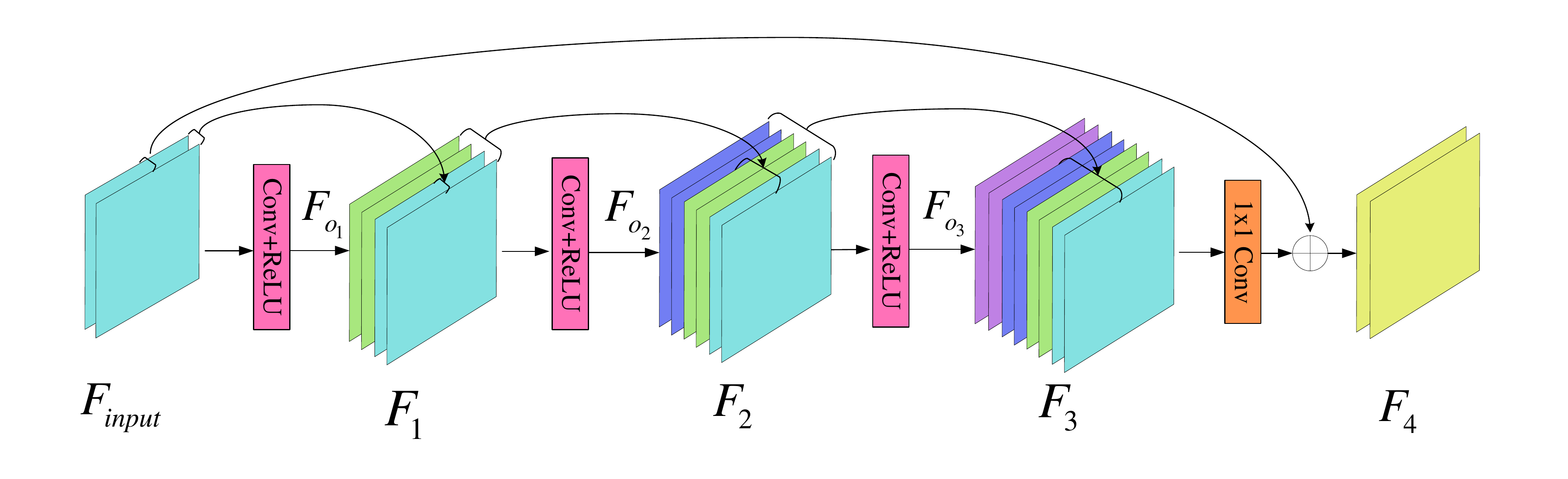}\\
% \caption{Architecture of FB as well as RDB.}
\caption{Architecture of feature enhancement (FE) as well as residual dense block (RDB).}
\end{figure}

\section{Experiments}
\subsection{Datasets}
\textbf{Training datasets:} Berkeley segmentation dataset (BSD) with 432 natural images of ${481 \times 321}$ \cite{li2013benchmark} to train color and gray Gaussian synthetic denoisers. To enhance the number of training images and accelerate training efficiency, we randomly cut each image into 512 image patches of ${48 \times 48}$. The number of noisy image patches is 221,184. In terms of real-noisy image denoising, we choose 100 natural images with sizes of ${512 \times 512}$ as training data to train a real noisy image denoiser \cite{xu2018real}.  Also, we cut each image in real noisy image dataset as 211,600 image patches of ${48 \times 48}$. To further extend diversity of training images, we randomly use a one way from data augmentation with eight ways to make training images richer, mentioned eight ways can be shown as Ref. \cite{tian2020attention}.

\textbf{Test datasets:} To fairly test the proposed denoising method, public datasets, i.e., BSD68 \cite{li2013benchmark}, Set12 \cite{li2013benchmark}, CBSD68 \cite{li2013benchmark}, Kodak24 \cite{franzen1999kodak} and CC \cite{nam2016holistic}. In terms of gray Gaussian synthetic image denoising, BSD68 \cite{li2013benchmark} and Set12 \cite{li2013benchmark} are used as test datasets. In terms of color Gaussian synthetic image denoising, CBSD68 \cite{li2013benchmark} and Kodak24 \cite{franzen1999kodak} are utilized as test datasets. In terms of real noisy image denoising, CC \cite{nam2016holistic} is exploited as test dataset.

\subsection{Implementation details}
 For training denoisers in this paper, batch size and the number of epochs are set to 64 and 90,  respectively. Learning rate is initialized as ${1 \times 10^{-4}}$, which may vary from different epochs. That is, learning rate is ${1 \times 10^{-4}}$ from the 1st epoch to the 30th epoch, it is ${1 \times 10^{-5}}$ from the 31th epoch to the 60th epoch and it is ${1 \times 10^{-6}}$ from the 61th epoch to the 90th epoch. ${{\beta _1} = 0.9}$ and ${\beta _2} = 0.999$.  More experimental parameters are the same as Ref. \cite{tian2020attention}. We use PyTorch of 1.10.2 \cite{tian2020deep} and Python of 3.8.5 to implement codes of MWDCNN. All the experiments are conducted on Ubuntu of 20.04 with AMD EPYC of 7502P/3.35GHz, 32-core CPU, RAM of 128G and a GPU of a Nvidia GeForce GTX 3090. Besides, we choose Nvidia CUDA of 11.1 version and cuDNN of 8.04 version to improve running speed of the mentioned GPU. 
 
\subsection{Network analysis}
In this paper, we analyze designed denoising network to show its rationality and validity as follows. 

Dynamic convolutional block:  Robust denoisers are very important for different scenes. However, most of existing denoising methods resorted to sharing parameters to train denoising models, which cannot make these denoising models robust for different scenes. That also has huge challenges for high-level vision tasks. Also, dynamic convolution can adjust parameters of convolutional kernels, according to given images \cite{chen2020dynamic}. Inspired by that, we fuse a dynamic convolution into a CNN to suppress noise for improving denoising performance without increasing computational costs. That is, we design a dynamic convolutional block as well as DCB to address the mentioned problems. DCB uses a convolutional layer to extract linear features. Subsequently, a dynamic convolutional layer \cite{chen2020dynamic} is used in a CNN to linearly combine all convolutional kernels to dynamically adjust parameters rather than same parameters of convolutions, according to different noisy images, which can make a tradeoff between denoising performance and computational costs as follows.

In terms of denoising performance, we design three experiments to verify denoising effect of a dynamic convolution in a CNN, according to individual and local parts  of the MWDCNN for image denoising. In terms of a dynamic convolution as a local part in a MWDCNN, we use MWDCNN without a convolution in a DCB and MWDCNN without a dynamic convolution and a convolutional layer in a DCB to verify superiority of a dynamic convolution for denoising effect as shown in Table 1, which shows effectiveness of dynamic convolutional block in the MWDCNN for image denoising. In terms of a dynamic convolution as an individual, we use combination of a dynamic convolution layer and three convolutional layers as well as the combination of three stacked convolutional layers and a dynamic convolutional layer and combination of six stacked convolutional layers (also treated as four convolutions with ${5 \times 5}$ and two convolutions with ${1 \times 1}$ ) to conduct experiments. The number of input and output channels of the first convolutional layer are 3 and 64, respectively. The number of input and output channels of the last convolutional layer are 64 and 3, respectively. The number of input and output channels of other convolutional layers are 64 and 64, respectively. In Table 1, we can see that the combination of three stacked convolutional layers and a dynamic convolutional layer obtains higher value than that of combination of six stacked convolutional layers, which tests effectiveness of individual for image denoising.
According to mentioned analysis, it is known that a dynamic convolution in a CNN is effective for image denoising.

In terms of computational costs, we choose important metrices, i.e., complexities (parameters and Flops) and run-time to show denoising competitiveness of a dynamic convolution on real digital cameras in the real world. That is, the combination of three stacked convolutional layers and a dynamic convolutional layer obtains competitive both parameters and flops than that of combination of six stacked convolutional layers as illustrated in Table 2, which shows competitive complexities of a dynamic convolution for image denoising. The combination of three stacked convolutional layers and a dynamic convolutional layer obtains  less run-time than that of combination of six stacked convolutional layers in Table 3, which shows fast execution denoising ability of a dynamic convolutional layer. Thus, a dynamic convolutional layer used in a CNN is very reasonable and valid for image denoising in this paper. Besides, we use a convolutional layer in a DCB to extract useful information, where its effectiveness is tested via MWDCNN and MWDCNN without a convolution in a DCB in Table 1. 

%%%%%%%%%%%%%%%%%%%%%%%%%%%%%%%%%%%%%%%%%%%%%%%%%%%%%%%%%%%%%%%%%%%%%%%%%%%%%%%%%%%%%%%%%%%%%%%%%%%%%%%%%%%%%%%%
%%%%%%%%%%%%%%%%%%%%%%%%%%%%%%%%%%%%%%%%%%%%%%%%%Table1%%%%%%%%%%%%%%%%%%%%%%%%%%%%%%%%%%%%%%%%%%%%%%%%%%%%%%%%%
%%%%%%%%%%%%%%%%%%%%%%%%%%%%%%%%%%%%%%%%%%%%%%%%%%%%%%%%%%%%%%%%%%%%%%%%%%%%%%%%%%%%%%%%%%%%%%%%%%%%%%%%%%%%%%%%

\begin{table}[htbp!]
\caption{PSNR(dB) of different methods on CBSD68 for ${\sigma = 25}$.}
\begin{tabular}{|l|c|lll}
\cline{1-2}
\multicolumn{1}{|c|}{Methods}                                                                                                                                       & PSNR  &  &  &  \\ \cline{1-2}
MWDCNN (Ours)                                                                                                                                                             & 31.45 &  &  &  \\ \cline{1-2}
MWDCNN without a convolution in a DCB                                                                                                                               & 31.42 &  &  &  \\ \cline{1-2}
MWDCNN without a dynamic convolution and a convolutional layer in a DCB                                                                                             & 31.39 &  &  &  \\ \cline{1-2}
MWDCNN with a convolutional layer and without a DCB and a WEB                                                                                                       &31.35 &  &  &  \\ \cline{1-2}
MWDCNN with a convolutional layer and FE without a DCB and a WEB                                                                                                    & 31.34 &  &  &  \\ \cline{1-2}
MWDCNN with a convolutional layer and without DCB and two WEBs                                                                                                      & 31.33 &  &  &  \\ \cline{1-2}
\begin{tabular}[c]{@{}l@{}}MWDCNN with a convolutional layer and without DCB, two WEBs and two ReLUs \\in RDC\end{tabular}                                         & 31.31 &  &  &  \\ \cline{1-2}
\begin{tabular}[c]{@{}l@{}}MWDCNN with a convolutional layer and without DCB, two WEBs, two ReLUs and \\two RLOs in RDC\end{tabular}                               & 31.30 &  &  &  \\ \cline{1-2}
\begin{tabular}[c]{@{}l@{}}MWDCNN with a convolutional layer and  without DCB, two WEBs, two ReLUs, \\two RLOs and a RDB in RDC\end{tabular}                       & 31.20 &  &  &  \\ \cline{1-2}
\begin{tabular}[c]{@{}l@{}}MWDCNN with a convolutional layer and without DCB, two WEBs, two ReLUs, \\two RLOs, a RDB and a convolutional layer in RDC\end{tabular} & 31.12 &  &  &  \\ \cline{1-2}
The combination of six stacked convolutional layers                                                                                                                 & 30.83 &  &  &  \\ \cline{1-2}
\begin{tabular}[c]{@{}l@{}}The combination of three stacked convolutional layers and a dynamic convolutional\\ layer\end{tabular}                                  & 30.85 &  &  &  \\ \cline{1-2}

\begin{tabular}[c]{@{}l@{}} MWDCNN without two WEBs \end{tabular}                                  & 31.35 &  &  &  \\ \cline{1-2}

\begin{tabular}[c]{@{}l@{}} MWDCNN with a convolutional layer and without an RDC\end{tabular}                                  & 31.32 &  &  &  \\ \cline{1-2}

\begin{tabular}[c]{@{}l@{}} MWDCNN without ReLUs\end{tabular}                                  & 31.44 &  &  &  \\ \cline{1-2}

\end{tabular}
\end{table}
\label{tab:booktabs}

%%%%%%%%%%%%%%%%%%%%%%%%%%%%%%%%%%%%%%%%%%%%%%%%%%%%%%%%%%%%%%%%%%%%%%%%%%%%%%%%%%%%%%%%%%%%%%%%%%%%%%%%%%%%%%%%
%%%%%%%%%%%%%%%%%%%%%%%%%%%%%%%%%%%%%%%%%%%%%%%%%Table2%%%%%%%%%%%%%%%%%%%%%%%%%%%%%%%%%%%%%%%%%%%%%%%%%%%%%%%%%
%%%%%%%%%%%%%%%%%%%%%%%%%%%%%%%%%%%%%%%%%%%%%%%%%%%%%%%%%%%%%%%%%%%%%%%%%%%%%%%%%%%%%%%%%%%%%%%%%%%%%%%%%%%%%%%%
\begin{table}[htbp!]
\caption{Complexities of different methods on CBSD68 for ${\sigma = 25}$.}
\centering
\begin{tabular}{|l|c|c|}
\hline
\multicolumn{1}{|c|}{Methods}                                                                                                      & Parameters & Flops \\ \hline
The combination of six stacked convolutional layers                                                                                & 0.212M     & 5.59G \\ \hline
\begin{tabular}[c]{@{}l@{}}The combination of three stacked convolutional layers\\  and a dynamic convolutional layer\end{tabular} & 0.498M     & 2.81G \\ \hline
\end{tabular}
\end{table}
\label{tab:booktabs}
%%%%%%%%%%%%%%%%%%%%%%%%%%%%%%%%%%%%%%%%%%%%%%%%%%%%%%%%%%%%%%%%%%%%%%%%%%%%%%%%%%%%%%%%%%%%%%%%%%%%%%%%%%%%%%%%
%%%%%%%%%%%%%%%%%%%%%%%%%%%%%%%%%%%%%%%%%%%%%%%%%Table3%%%%%%%%%%%%%%%%%%%%%%%%%%%%%%%%%%%%%%%%%%%%%%%%%%%%%%%%%
%%%%%%%%%%%%%%%%%%%%%%%%%%%%%%%%%%%%%%%%%%%%%%%%%%%%%%%%%%%%%%%%%%%%%%%%%%%%%%%%%%%%%%%%%%%%%%%%%%%%%%%%%%%%%%%%

\begin{table}[htbp!]
\caption{Run-time of different methods on a noisy image of $ {1024 \times 1024}$ from the CBSD68 for ${\sigma = 25}$.}
\centering
\begin{tabular}{|l|c|c|}
\hline
\multicolumn{1}{|c|}{Methods}                                                                                                      & Run-time  \\ \hline
The combination of six stacked convolutional layers                                                                                & 0.060s    \\ \hline
\begin{tabular}[c]{@{}l@{}}The combination of three stacked convolutional layers\\  and a dynamic convolutional layer\end{tabular} & 0.046s    \\ \hline
\end{tabular}
\end{table}
\label{tab:booktabs}

%%%%%%%%%%%%%%%%%%%%%%%%%%%%%%%%%%%%%%%%%%%%%%%%%%%%%%%%%%%%%%%%%%%%%%%%%%%%%%%%%%%%%%%%%%%%%%%%%%% %%%%%%%%%%%%%%%%%%%%%%%%%%%%%%%%%%%%%%%%%%%%Table4%%%%%%%%%%%%%%%%%%%%%%%%%%%%%%%%%%%%%%%%%%%%%%%%% %%%%%%%%%%%%%%%%%%%%%%%%%%%%%%%%%%%%%%%%%%%%%%%%%%%%%%%%%%%%%%%%%%%%%%%%%%%%%%%%%%%%%%%%%%%%%%%%%%%

% \usepackage{color}

\begin{table}
\caption{PSNR(dB) and SSIM of MWDCNN with different loss functions on CBSD68 for $\sigma=25$.}
\centering
\begin{tabular}{|l|l|l|} 
\hline
Loss function &PSNR  &SSIM    \\ 
\hline
MSE\cite{allen1971mean}           & 31.45 & 0.8925  \\ 
\hline
Charbonnier\cite{charbonnier1994two}   & 31.43 & 0.8918  \\ 
\hline
MSE+Pearson\cite{ayyoubzadeh2021asna}   & 31.42 & 0.8919  \\
\hline
\end{tabular}
\end{table}

Wavelet transform enhancement block: It is known that images can be treated as signals, thus, signal processing techniques, i.e., wavelet transform are effective for low-level tasks \cite{cho2005multivariate}. Besides, a combination of a CNN and wavelet transform is very effective for low-level vision tasks in Section 2.3. Inspired by that, we fuse wavelet transform techniques into a CNN for image denoising. Differing from these methods in Section 2.3, we embed a network architecture into a wavelet transform technique to fuse frequency features and structure information via a feature enhancement mechanism and DWT as shown in Section 3.4, where effectiveness of wavelet transform technique can be tested via MWDCNN with a convolutional layer and without a DCB and a WEB and MWDCNN with a convolutional layer and FE without a DCB and a WEB in Table 1. That shows effectiveness of signal processing technique (wavelet technique) in the MWDCNN for image denoising. MWDCNN with a convolutional layer and FE without a DCB and a WEB has higher PSNR value than that of MWDCNN with a convolutional layer and without DCB and two WEBs, which shows the effectiveness of FE in image denoising in Table 1.That shows effectiveness of FE (discriminative learning) in a MWDCNN for image denoising. To extract more robust features, we choose two stacked WEBs in a CNN in image denoising in Section 3.4. Its effectiveness is verified via MWDCNN without a dynamic convolution and a convolutional layer in a DCB and MWDCNN with a convolutional layer and without a DCB and a WEB, MWDCNN with a convolutional layer and without a DCB and a WEB and MWDCNN with a convolutional layer and without DCB and two WEBs as listed in Table 1. Besides, MWDCNN is superior to MWDCNN without two WEBs, which tests effectiveness of the combination of signal processing and discriminative learning techniques in the MWDCNN for image denoising in Table 1. According to mentioned illustrations, two stacked WEBs in a MWDCNN are effective and reasonable for image denoising. 

Residual block: According to Ref. \cite{ahn2018image}, we can see that stacked convolutional layers can refine obtained features. To enhance robustness of obtained denoiser and prevent dependency of deep networks, we propose a 10-layer residual block as well as RB via improved residual dense architectures to improve quality of predicted images and construct images as follows. Firstly, we use two stacked RDBs (each RDB with three convolution kernels of ${5 \times 5}$) to refine obtained features from stacked WEBs, its effectiveness is tested by MWDCNN with a convolutional layer and without DCB, two WEBs, two ReLUs and two RLOs in RDC and MWDCNN with a convolutional layer and without DCB, two WEBs, two ReLUs, two RLOs and a RDB in RDC in Table 1. Secondly, ReLU acts each RDB to convert obtained features into non-linearity, where its effectiveness is verified through MWDCNN with a convolutional layer and without DCB and two WEBs and MWDCNN with a convolutional layer and without DCB, two WEBs and two ReLUs in RDC as presented in Table 1. To address long-term dependency of deep networks, we use two residual operations to fuse obtained features from the second combination of RDB and ReLU and the second WEB. MWDCNN with a convolutional layer and without DCB, two WEBs and two ReLUs in RDC outperforms MWDCNN with a convolutional layer and without DCB, two WEBs, two ReLUs and two RLOs in RDC in PSNR as listed in Table 1, which shows its effectiveness. Fourthly, we use a convolutional layer of $5\times5$  to prevent over-enhancement phenomenon from the previous steps. MWDCNN with a convolutional layer and without DCB, two WEBs, two ReLUs, two RLOs and a RDB in RDC has obtained higher PSNR value than that of MWDCNN with a convolutional layer and without DCB, two WEBs, two ReLUs, two RLOs, a RDB and a convolutional layer in RDC to test effectiveness of a convolutional layer. According to mention illustrations, the designed block is reasonable and valid for image denoising. Also, MWDCNN has obtained better denoising performance than that of MWDCNN with a convolutional layer and without an RDC for image denoising in Table 1, which verifies effectiveness of RDC for MWDCNN in image denoising. Besides, ReLUs as important  components of MWDCNN  are used to convert linear data into non-linearity, where its effectiveness is verified by MWDCNN and MWDCNN without ReLUs in Table 1.

In terms of choosing loss function, we take into the following three aspects consideration. The first aspect utilizes a loss function as well as MSE of popular denoising methods as an objective function in terms of previous denoising work. The second aspect uses a loss function (Charbonnier) of the newest denoising method based Transformer to fuse into the proposed MWDCNN to test denoising performance in terms of the newest denoising work. The third aspect exploits typical quality assessment method of Pearson as a loss function to test denoising performance in terms of perceptual theory. As shown in Table 4, we can see that the proposed method with MSE is superior to the proposed method with Charbonnier, which shows MSE is suitable to lightweight denoising CNN and Charbonnier \cite{liang2021swinir} is useful for big denoising CNN, i.e., Transformer. Also, only using MSE as a loss function outperforms using the combination of MSE and Pearson as a loss function \cite{ayyoubzadeh2021asna} in terms of PSNR and SSIM for image denoising, which shows fusion of two different losses may cause a negative effect. Thus, we choose MSE as a loss function to train a MWDCNN denoiser.

In a summary, the dynamic convolutional block with a dynamic convolution is strong adaptability for different scenes. Also, it is very competitive in terms of denoising performance and computational costs. To further improve denoising performance, a wavelet transform enhancement block relies on  a combination of a single processing technique and discriminative learning technique via fusing wavelet transform technique and a CNN is designed. To improve the robustness of obtained MWDCNN denoising model, a residual block uses a stacked architecture to refine obtained information. According to mentioned illusrations, we can see that key  components, i.e.,  dynamic convolutional block, wavelet transform enhancement block, residual block and chosen loss function are reasonable and valid for image denoising, which makes MWDCNN reasonable and valid in image denoising in this paper.

% \usepackage{color}

%%%%%%%%%%%%%%%%%%%%%%%%%%%%%%%%%%%%%%%%%%%%%%%%%%%%%%%%%%%%%%%%%%%%%%%%%%%%%%%%%%%%%%%%%%%%%%%%%%%%%%%%%%%%%%%%
%%%%%%%%%%%%%%%%%%%%%%%%%%%%%%%%%%%%%%%%%%%%%%%%Table4%%%%%%%%%%%%%%%%%%%%%%%%%%%%%%%%%%%%%%%%%%%%%%%%%%%%%%%%%%
%%%%%%%%%%%%%%%%%%%%%%%%%%%%%%%%%%%%%%%%%%%%%%%%%%%%%%%%%%%%%%%%%%%%%%%%%%%%%%%%%%%%%%%%%%%%%%%%%%%%%%%%%%%%%%%%
\begin{table}[htbp!]
\centering
\caption{Average PSNR (dB) of different methods on BSD68 with noise levels of 15, 25 and 50.}
\begin{tabular}{|l|c|c|c|} 
\hline
\multicolumn{1}{|c|}{Methods} & \begin{tabular}[c]{@{}c@{}}15\\\end{tabular}                                    & \begin{tabular}[c]{@{}c@{}}25\\\end{tabular}                     & \begin{tabular}[c]{@{}c@{}}50\\\end{tabular}                                     \\ 
\hline
BM3D\cite{dabov2007image}                      & 31.07                                                                           & 28.57                                                            & 25.62                                                                            \\ 
\hline
WNNM\cite{gu2014weighted}                      & 31.37                                                                           & 28.83                                                            & 25.87                                                                            \\ 
\hline
TNRD\cite{chen2016trainable}                      & 31.42                                                                           & 28.92                                                            & 25.97                                                                            \\ 
\hline
DnCNN\cite{zhang2017beyond}                     & 31.72                                                                           & 29.23                                                            & 26.23                                                                            \\ 
\hline
FFDNet\cite{zhang2018ffdnet}                    & 31.63                                                                           & 29.19                                                            & \textcolor{blue}{26.29}                                                                            \\ 
\hline
ADNet\cite{tian2020attention}                     & \begin{tabular}[c]{@{}c@{}}\textcolor{blue}{31.74}\\\end{tabular} & \textcolor{blue}{29.25}                                                            & \textcolor{blue}{26.29}                                                                            \\ 
\hline
CDNet\cite{quan2021image}                     & \begin{tabular}[c]{@{}c@{}}\textcolor{blue}{31.74}\\\end{tabular} & \begin{tabular}[c]{@{}c@{}}\textcolor{red}{29.28}\\\end{tabular} & \begin{tabular}[c]{@{}c@{}}\textcolor{red}{26.36}\\\end{tabular}                 \\ 
\hline
MWDCNN (Ours)                  & \begin{tabular}[c]{@{}c@{}}\textcolor{red}{31.77}\\\end{tabular}                & \begin{tabular}[c]{@{}c@{}}\textcolor{red}{29.28}\\\end{tabular} & \textcolor{blue}{26.29}                                                                            \\ 
\hline
MWDCNN-B (Ours)                & 31.39                                                                           & 29.16                                                            & 26.20
\\
\hline
\end{tabular}
\end{table}
\label{tab:booktabs}

%%%%%%%%%%%%%%%%%%%%%%%%%%%%%%%%%%%%%%%%%%%%%%%%%%%%%%%%%%%%%%%%%%%%%%%%%%%%%%%%%%%%%%%%%%%%%%%%%%%%%%%%%%%%%%%%
%%%%%%%%%%%%%%%%%%%%%%%%%%%%%%%%%%%%%%%%%%%%%%%%%Table5%%%%%%%%%%%%%%%%%%%%%%%%%%%%%%%%%%%%%%%%%%%%%%%%%%%%%%%%%
%%%%%%%%%%%%%%%%%%%%%%%%%%%%%%%%%%%%%%%%%%%%%%%%%%%%%%%%%%%%%%%%%%%%%%%%%%%%%%%%%%%%%%%%%%%%%%%%%%%%%%%%%%%%%%%%
% \usepackage{color}

\begin{table}[htbp!]
\centering
\caption{Average PSNR (dB) of different methods on Set12 with noise levels of 15, 25 and 50.}
\tiny
\begin{tabular}{|l|c|c|c|c|c|c|c|c|c|c|c|c|c|} 
\hline
\multicolumn{1}{|c|}{Images}      & C.man                                  & House                                  & Peppers                               & Starfish                              & Monarch                                    & Airplane                               & Parrot                                & Lena                                  & Barbara                               & Boat                                  & Man                                   & Couple                                & Average                                \\ 
\hline
\multicolumn{1}{|c|}{noise level} & \multicolumn{13}{c|}{15}                                                                                                                                                                                                                                                                                                                                                                                                                                                                                                                       \\ 
\hline
BM3D\cite{dabov2007image}                          & 31.91                                  & 34.93                                  & 32.69                                 & 31.14                                 & 31.85                                      & 31.07                                  & 31.37                                 & 34.26                                 & \textcolor{blue}{33.10} & 32.13                                 & 31.92                                 & 32.10                                 & 32.37                                  \\ 
\hline
WNNM\cite{gu2014weighted}                          & 32.17                                  & \textcolor{red}{35.13 } & 32.99                                 & 31.82                                 & 32.71                                      & 31.39                                  & 31.62                                 & 34.27                                 & \textcolor{red}{33.60}                & 32.27                                 & 32.11                                 & 32.17                                 & 32.70                                  \\ 

\hline
TNRD\cite{chen2016trainable}                          & 32.19                                  & 34.53                                  & 33.04                                 & 31.75                                 & 32.56                                      & 31.46                                  & 31.63                                 & 34.24                                 & 32.13                                 & 32.14                                 & 32.23                                 & 32.11                                 & 32.50                                  \\ 
\hline
DnCNN\cite{zhang2017beyond}                         & \textcolor{red}{32.61}                 & 34.97                                  & \textcolor{red}{33.30}                & \textcolor{blue}{32.20}                                 & \textcolor{blue}{33.09}                                      & \textcolor{blue}{31.70}  & \textcolor{blue}{31.83}                                 & \textcolor{blue}{34.62} & 32.64                                 & \textcolor{blue}{32.42} & \textcolor{red}{32.46}                & \textcolor{blue}{32.47} & \textcolor{blue}{32.86}  \\ 
\hline
FFDNet\cite{zhang2018ffdnet}                        & 32.43                                  & 35.07                                  & 33.25                                 & 31.99                                 & 32.66                                      & 31.57                                  & 31.81                                 & \textcolor{blue}{34.62} & 32.54                                 & 32.38                                 & \textcolor{blue}{32.41}                                 & 32.46                                 & 32.77                                  \\ 
\hline
GNLM\cite{li2016novel}                         &  {——}                                  & 35.01                & 32.98                                 & {——} & {——}                    &  {——}                                  & {——} & 33.89                                 & {——}   & 31.69  & 31.94               & {——}                                 & {——}  \\ 
\hline
MWDCNN(Ours)                      & \textcolor{blue}{32.53}                                  & \textcolor{blue}{35.09}                                  & \textcolor{blue}{33.29} & \textcolor{red}{32.28}                & \textcolor{red}{33.20}                     & \textcolor{red}{31.74}                 & \textcolor{red}{31.97}                & \textcolor{red}{34.64}                & 32.65                                 & \textcolor{red}{32.49}                & \textcolor{red}{32.46}                & \textcolor{red}{32.52}                & \textcolor{red}{32.91}                 \\ 
\hline
MWDCNN-B(Ours)                    & 32.14                                  & 34.98                                  & 33.17                                 & 31.73                                 & 33.03                                      & 31.49                                  & 31.78                                 & 34.46                                 & 31.79                                 & 32.22                                 & 32.17                                 & 32.26                                 & 32.60                                  \\ 
\hline
\multicolumn{1}{|c|}{noise level} & \multicolumn{13}{c|}{25}                                                                                                                                                                                                                                                                                                                                                                                                                                                                                                                       \\ 
\hline
BM3D\cite{dabov2007image}                          & 29.45                                  & 32.85                                  & 30.16                                 & 28.56                                 & 29.25                                      & 28.42                                  & 28.93                                 & 32.07                                 & \textcolor{blue}{30.71} & 29.90                                 & 29.61                                 & 29.71                                 & 29.97                                  \\ 
\hline
WNNM\cite{gu2014weighted}                          & 29.64                                  & 33.22                                  & 30.42                                 & 29.03                                 & 29.84                                      & 28.69                                  & 29.15                                 & 32.24                                 & \textcolor{red}{31.24}                & 30.03                                 & 29.76                                 & 29.82                                 & 30.26                                  \\ 
\hline
TNRD\cite{chen2016trainable}                          & 29.72                                  & 32.53                                  & 30.57                                 & 29.02                                 & 29.85                                      & 28.88                                  & 29.18                                 & 32.00                                 & 29.41                                 & 29.91                                 & 29.87                                 & 29.71                                 & 30.06                                  \\ 
\hline
DnCNN\cite{zhang2017beyond}                         & 30.18                                  & 33.06                                  & 30.87 & 29.41                                 & 30.28                                      & 29.13                                  & 29.43                                 & 32.44                                 & 30.00                                 & 30.21                                 & \textcolor{blue}{30.10} & 30.12                                 & 30.43                                  \\ 
\hline
FFDNet\cite{zhang2018ffdnet}                        & 30.10                                  & \textcolor{blue}{33.28}  & \textcolor{blue}{30.93}                                 & 29.32                                 & 30.08                                      & 29.04                                  & \textcolor{blue}{29.44}                                 & \textcolor{blue}{32.57} & 30.01                                 & \textcolor{blue}{30.25} & \textcolor{red}{30.11}                & \textcolor{red}{30.20}                                 & 30.44                                  \\ 
\hline
DudeNet\cite{tian2021designing}                       & \textcolor{red}{30.23}                 & 33.24                                  & \textcolor{red}{30.98}                & \textcolor{blue}{29.53} & \textcolor{blue}{30.44}      & \textcolor{blue}{29.14}  & \textcolor{red}{29.48}                & 32.52                                 & 30.15                                 & 30.24                                 & 30.08                                 & \textcolor{blue}{30.15}                & \textcolor{blue}{30.52}                 \\ 
\hline
GNLM\cite{li2016novel}                         &  {——}                                  & 32.91                & 30.19                                 & {——} & {——}                    &  {——}                                  & {——} & 31.67                                 & {——}   & 29.71  & 29.63               & {——}                                 & {——}  \\
\hline
MWDCNN(Ours)                      & \textcolor{blue}{30.19}  & \textcolor{red}{33.33}                 & 30.85                                 & \textcolor{red}{29.66}                & \textcolor{red}{30.55}                     & \textcolor{red}{29.16}                 & \textcolor{red}{29.48}                & \textcolor{red}{32.67}                & 30.21                                 & \textcolor{red}{30.28}                & \textcolor{blue}{30.10} & 30.13                                 & \textcolor{red}{30.55}                 \\ 
\hline
MWDCNN-B(Ours)                    & 30.02                                  & 33.14                                  & 30.75                                 & 29.29                                 & 30.28                                      & 29.02                                  & 29.36                                 & 32.51                                 & 29.90                                 & 30.17                                 & 30.05                                 & 30.14 & 30.39                                  \\ 
\hline
\multicolumn{1}{|c|}{noise level} & \multicolumn{13}{c|}{50}                                                                                                                                                                                                                                                                                                                                                                                                                                                                                                                       \\ 
\hline
BM3D\cite{dabov2007image}                          & 26.13                                  & 29.69                                  & 26.68                                 & 25.04                                 & 25.82                                      & 25.10                                  & 25.90                                 & 29.05                                 & \textcolor{blue}{27.22} & 26.78                                 & 26.81                                 & 26.46                                 & 26.72                                  \\ 
\hline
WNNM\cite{gu2014weighted}                          & 26.45                                  & 30.33                                  & 26.95                                 & 25.44                                 & 26.32                                      & 25.42                                  & 26.14                                 & 29.25                                 & \textcolor{red}{27.79}                & 26.97                                 & 26.94                                 & 26.64                                 & 27.05                                  \\ 
\hline
TNRD\cite{chen2016trainable}                          & 26.62                                  & 29.48                                  & 27.10                                 & 25.42                                 & 26.31                                      & 25.59                                  & 26.16                                 & 28.93                                 & 25.70                                 & 26.94                                 & 26.98                                 & 26.50                                 & 26.81                                  \\ 
\hline
DnCNN\cite{zhang2017beyond}                         & 27.03                                  & 30.00                                  & 27.32                                 & 25.70                                 & 26.78                                      & 25.87                                  & 26.48                                 & 29.39                                 & 26.22                                 & 27.20                                 & 27.24                                 & 26.90                                 & 27.18                                  \\ 
\hline
FFDNet\cite{zhang2018ffdnet}                        & 27.05                                  & \textcolor{blue}{30.37}  & \textcolor{red}{27.54}                & 25.75                                 & 26.81                                      & \textcolor{blue}{25.89}  & \textcolor{red}{26.57}                & \textcolor{red}{29.66}                & 26.45                                 & \textcolor{red}{27.33}                & \textcolor{red}{27.29}                & \textcolor{blue}{27.08} & \textcolor{blue}{27.32}  \\ 
\hline
DudeNet\cite{tian2021designing}                       & \textcolor{red}{27.22}                 & 30.27                                  & \textcolor{blue}{27.51} & \textcolor{red}{25.88}                & \textcolor{blue}{26.93}      & 25.88                                  & \textcolor{blue}{26.50} & 29.45                                 & 26.49                                 & \textcolor{blue}{27.26}                                 & 27.19                                 & 26.97                                 & 27.30                                  \\ 
\hline
GNLM\cite{li2016novel}                         &  {——}                                  & 28.99                & 26.96                                 & {——} & {——}                    &  {——}                                  & {——} & 28.49                                 & {——}   & 26.63  & 26.78               & {——}                                 & {——}  \\
\hline
MWDCNN(Ours)                      & 26.99                                  & \textcolor{red}{30.58}                 & 27.34                                 & \textcolor{blue}{25.85} & \textcolor{red}{27.02}                     & \textcolor{red}{25.93}                 & 26.48                                 & \textcolor{blue}{29.63} & 26.60                                 & 27.23                                 & \textcolor{blue}{27.27} & \textcolor{red}{27.11}                & \textcolor{red}{27.34}                 \\ 
\hline
MWDCNN-B(Ours)                    & \textcolor{blue}{27.09}                                  & 30.23                                  & 27.35                                 & 25.70                                 & 26.78                                      & 25.73                                  & 26.39                                 & 29.42                                 & 26.57                                 & 27.24 & 27.23                                 & 26.97                                 & 27.23                                  \\ 
\hline
\end{tabular}
\end{table}
\label{tab:booktabs}
%%%%%%%%%%%%%%%%%%%%%%%%%%%%%%%%%%%%%%%%%%%%%%%%%%%%%%Table6%%%%%%%%%%%%%%%%%%%%%%%%%%%%%%%%%%%%%%%%%%%%%%%%%%
% \usepackage{color}

\begin{table}[htbp!]
\centering
\caption{Average PSNR (dB) of different methods on CBSD68 with noise levels of 15, 25, 35, 50 and 75.}
\begin{tabular}{|l|c|c|c|c|c|} 
\hline
\multicolumn{1}{|c|}{Methods} & 15                                    & 25                                    & 35                                    & 50                                    & 75                                     \\ 
\hline
CBM3D\cite{dabov2007image}                     & 33.52                                 & 30.71                                 & 28.89                                 & 27.38                                 & 25.74                                  \\ 
\hline
DnCNN\cite{zhang2017beyond}                     & 33.98                                 & 31.31                                 & 29.65                                 & 28.01                                 & ——                                     \\ 
\hline
FFDNet\cite{zhang2018ffdnet}                    & 33.80                                 & 31.18                                 & 29.57                                 & 27.96                                 & 26.24                                  \\ 
\hline
ADNet\cite{tian2020attention}                     & 33.99                                 & 31.31                                 & 29.66                                 & 28.04                                 & \textcolor{blue}{26.33}                                  \\ 
\hline
DudeNet\cite{tian2021designing}                   & 34.01                                 & 31.34                                 & 29.71                                 & 28.09                                 & \textcolor{red}{26.40}  \\ 
\hline
GradNet\cite{liu2020gradnet}                   & 34.07                                 & 31.39                                 & ——                                    & 28.12                                 & ——                                     \\ 
\hline
MWDCNN (Ours)                 & \textcolor{red}{34.18}                & \textcolor{red}{31.45}                & \textcolor{red}{29.81}                & \textcolor{blue}{28.13}                                 & \textcolor{red}{26.40}  \\ 
\hline
MWDCNN-B (Ours)                & \textcolor{blue}{34.10} & \textcolor{blue}{31.44} & \textcolor{blue}{29.80} & \textcolor{red}{28.15} & ——                                     \\ 
\hline
\end{tabular}
\end{table}
\label{tab:booktabs}
%%%%%%%%%%%%%%%%%%%%%%%%%%%%%%%%%%%%%%%%%%%%%%%%%%%%%%%%%%%%%%%%%%%%%%%%%%%%%%%%%%%%%%%%%%%%%%%%%%%
%%%%%%%%%%%%%%%%%%%%%%%%%%%%%%%%%%%%%%%%%%%%Table7%%%%%%%%%%%%%%%%%%%%%%%%%%%%%%%%%%%%%%%%%%%%%%%%%
%%%%%%%%%%%%%%%%%%%%%%%%%%%%%%%%%%%%%%%%%%%%%%%%%%%%%%%%%%%%%%%%%%%%%%%%%%%%%%%%%%%%%%%%%%%%%%%%%%%
% \usepackage{color}
% \usepackage[normalem]{ulem}

\begin{table}[htbp!]
\centering
\caption{Average PSNR (dB) of different methods on Kodak24 with noise levels of 15, 25, 35, 50 and 75.}
\begin{tabular}{|l|c|c|c|c|c|} 
\hline
\multicolumn{1}{|c|}{Methods} & $\sigma=15$                           & $\sigma=25$                                   & $\sigma=35$                           & $\sigma=50$                           & $\sigma=75$                            \\ 
\hline
CBM3D\cite{dabov2007image}                     & 34.28                                 & 31.68                                         & 29.90                                 & 28.46                                 & 26.82                                  \\ 
\hline
DnCNN\cite{zhang2017beyond}                     & 34.73                                 & 32.23                                         & 30.64                                 & 29.02                                 & ——                                     \\ 
\hline
FFDNet\cite{zhang2018ffdnet}                    & 34.55                                 & 32.11                                         & 30.56                                 & 28.99                                 & 27.25                                  \\ 
\hline
ADNet\cite{tian2020attention}                     & 34.76                                 & 32.26                                         & 30.68                                 & 29.10                                 & \textcolor{blue}{27.40}                                  \\ 
\hline
DudeNet\cite{gu2014weighted}                   & 34.81                                 & 32.26                                         & 30.69                                 & 29.10                                 & 27.39                                  \\ 
\hline
GradNet\cite{liu2020gradnet}                   & \textcolor{blue}{34.85}                                 & 32.35                                         & ——                                    & 29.23 & ——                                     \\ 
\hline
MWDCNN (Ours)                 & \textcolor{red}{34.91}                & \textcolor{red}{32.40}                        & \textcolor{red}{30.87}                & \textcolor{red}{29.26}                & \textcolor{red}{27.55}                 \\ 
\hline
MWDCNN-B (Ours)                & 34.83                                 & \textcolor{blue}{32.39}                                         & \textcolor{blue}{30.83} & \textcolor{blue}{29.23} & ——                                     \\ 
\hline
\end{tabular}
\end{table}
\label{tab:booktabs}
%%%%%%%%%%%%%%%%%%%%%%%%%%%%%%%%%%%%%%%%%%%%%%%%%%%%%%%%%%%%%%%%%%%%%%%%%%%%%%%%%%%%%%%%%%%%%%%%%%% %%%%%%%%%%%%%%%%%%%%%%%%%%%%%%%%%%%%%%%%%%%%Table8%%%%%%%%%%%%%%%%%%%%%%%%%%%%%%%%%%%%%%%%%%%%%%%%%
%%%%%%%%%%%%%%%%%%%%%%%%%%%%%%%%%%%%%%%%%%%%%%%%%%%%%%%%%%%%%%%%%%%%%%%%%%%%%%%%%%%%%%%%%%%%%%%%%%%

% \usepackage{color}
% \usepackage[normalem]{ulem}
% \usepackage{multirow}

% \caption{PSNR (dB) of different methods on real noisy images.}
% \usepackage{color}
% \usepackage[normalem]{ulem}

\begin{table}
\caption{PSNR (dB) of different methods on real noisy images(CC).}
\centering
\begin{tabular}{|l|c|c|c|c|c|} 
\hline
Setting                                & CBM3D\cite{dabov2007image}                                                  & TID\cite{luo2015adaptive}   & DnCNN\cite{zhang2017beyond}                                & DudeNet\cite{tian2021designing}                                                & \begin{tabular}[c]{@{}c@{}}MWDCNN\\ (Ours)\textbf{}\end{tabular}  \\ 
\hline
\multirow{3}{*}{Canon 5D ISO = 3200}   & \textcolor{red}{39.76}                                 & 37.22   & \textcolor{blue}{37.26}                                 & 36.66                                  & 36.97                                                             \\ 
\cline{2-6}
                                       & \textcolor{blue}{36.40}                  & 34.54 & 34.13                                 & \textcolor{red}{36.70}                                 & 36.01\textcolor{red}{}                                            \\ 
\cline{2-6}
                                       & \textcolor{red}{36.37}                                 & 34.25 & 34.09                                 & \textcolor{blue}{35.03}                  & 34.80                                                     \\ 
\hline
\multirow{3}{*}{Nikon D600 ISO = 3200} & \textcolor{red}{34.18}                                 & 32.99 & 33.62                                 & 33.72                                                  & \textcolor{blue}{33.91}                             \\ 
\cline{2-6}
                                       & \textcolor{red}{35.07}\textcolor{blue}{} & 34.20 & 34.48                                 & 34.70                                          & \textcolor{blue}{34.88}                             \\ 
\cline{2-6}
                                       & \textcolor{blue}{37.13}                  & 35.58 & 35.41                                 & \textcolor{red}{37.98}\textcolor{blue}{} & 37.02                                                     \\ 
\hline
\multirow{3}{*}{Nikon D600 ISO = 3200} & 36.81                                                  & 34.49 & \textcolor{blue}{37.95} & \textcolor{red}{38.10}                                 & 37.93                                                             \\ 
\cline{2-6}
                                       & \textcolor{blue}{37.76}                  & 35.19 & 36.08                                 & \textcolor{red}{39.15}                                 & 37.49                                                             \\ 
\cline{2-6}
                                       & \textcolor{blue}{37.51}                  & 35.26 & 35.48                                 & 36.14                                                  & \textcolor{red}{38.44}                                            \\ 
\hline
\multirow{3}{*}{Nikon D600 ISO = 3200} & 35.05                                                  & 33.70 & 34.08                                 & \textcolor{blue}{36.93}          & \textcolor{red}{37.10}                                            \\ 
\cline{2-6}
                                       & 34.07                                                  & 31.04 & 33.70                                 & \textcolor{blue}{35.80}                  & \textcolor{red}{36.72}                                    \\ 
\cline{2-6}
                                       & 34.42                                                  & 33.07 & 33.31                                 & \textcolor{red}{37.49}                                 & \textcolor{blue}{37.25}                     \\ 
\hline
\multirow{3}{*}{Nikon D600 ISO = 3200} & 31.13                                                  & 29.40 & 29.83                                 & \textcolor{blue}{31.94}                  & \textcolor{red}{32.24}                                            \\ 
\cline{2-6}                                       & 31.22                                                  & 29.86 & 30.55                                 & \textcolor{blue}{32.51}          & \textcolor{red}{32.56}                                            \\ 
\cline{2-6}
                                       & 30.97                                                  & 29.21 & 30.09                                 & \textcolor{red}{32.91}                                 & \textcolor{blue}{32.76}                     \\ 
\hline
Average                                & 35.19                                                  & 33.36 & 33.86                                 & \textcolor{blue}{35.72}                  & \textcolor{red}{35.74}                                    \\
\hline
\end{tabular}
\end{table}

%%%%%%%%%%%%%%%%%%%%%%%%%%%%%%%%%%%

%%%%%%%%%%%%%%%%%%%%%%%%%%%%%%%%%%%%%%%%%%%%%%%%%%%%%%%%%%%%%%%%%%%%%%%%%%%%%%%%%%%%%%%%%%%%%%%%%%% %%%%%%%%%%%%%%%%%%%%%%%%%%%%%%%%%%%%%%%%%%%%Table9%%%%%%%%%%%%%%%%%%%%%%%%%%%%%%%%%%%%%%%%%%%%%%%%%
%%%%%%%%%%%%%%%%%%%%%%%%%%%%%%%%%%%%%%%%%%%%%%%%%%%%%%%%%%%%%%%%%%%%%%%%%%%%%%%%%%%%%%%%%%%%%%%%%%%
% \usepackage{color}

\begin{table}
\centering
\caption{FSIM results of different methods on BSD68 with noise levels of 15, 25 and 50.}
\begin{tabular}{|l|l|l|l|} 
\hline
Methods    & $\sigma=15$            & $\sigma=25$            & $\sigma=50$             \\ 
\hline
DnCNN\cite{zhang2017beyond}  & \textcolor{blue}{0.746}                  & 0.689                  & \textcolor{blue}{0.602}                   \\ 
\hline
FFDNet\cite{zhang2018ffdnet} & 0.745                  & \textcolor{blue}{0.690}                  & \textcolor{blue}{0.602}                   \\
\hline
MWDCNN (Ours)     & \textcolor{red}{0.747} & \textcolor{red}{0.691} & \textcolor{red}{0.603}  \\ 
\hline
\end{tabular}
\end{table}
\label{tab:booktabs}
%%%%%%%%%%%%%%%%%%%%%%%%%%%%%%%%%%%%%%%%%%%%%%%%%%%%%%%%%%%%%%%%%%%%%%%%%%%%%%%%%%%%%%%%%%%%%%%%%%% %%%%%%%%%%%%%%%%%%%%%%%%%%%%%%%%%%%%%%%%%%%%Table10%%%%%%%%%%%%%%%%%%%%%%%%%%%%%%%%%%%%%%%%%%%%%%%%% %%%%%%%%%%%%%%%%%%%%%%%%%%%%%%%%%%%%%%%%%%%%%%%%%%%%%%%%%%%%%%%%%%%%%%%%%%%%%%%%%%%%%%%%%%%%%%%%%%%
% \usepackage{color}

\begin{table}
\centering
\caption{FSIM results of different methods on CBSD68 and Kodak24 with noise levels of 15, 25, 35, 50 and 75.}
\begin{tabular}{|l|c|c|c|c|c|} 
\hline
\multicolumn{1}{|c|}{Noise Level} & $\sigma=15$\textcolor{red}{} & $\sigma=25$\textcolor{red}{} & $\sigma=35$\textcolor{red}{} & $\sigma=50$            & $\sigma=75$ \textcolor{red}{}  \\ 
\hline
\multicolumn{1}{|c|}{Dataset}     & \multicolumn{5}{c|}{CBSD68\textcolor{red}{}}                                                                                                         \\ 
\hline
DnCNN-B\cite{zhang2017beyond}                       & 0.784                        & 0.737                        & 0.701                        & 0.657                  & ——                             \\ 
\hline
ADNet\cite{tian2020attention}                         & \textcolor{blue}{0.785}                        & \textcolor{blue}{0.738}                        & \textcolor{blue}{0.702}                        & \textcolor{blue}{0.659}                  & \textcolor{blue}{0.606}                          \\ 
\hline
FFDNet\cite{zhang2018ffdnet}                        & 0.782                        & 0.733                        & 0.695                        & 0.648                  & 0.593                          \\ 
\hline
MWDCNN (Ours)                            & \textcolor{red}{0.788}       & \textcolor{red}{0.742}       & \textcolor{red}{0.707}       & \textcolor{red}{0.664} & \textcolor{red}{0.615}         \\ 
\hline
\multicolumn{1}{|c|}{Dataset}     & \multicolumn{5}{c|}{Kodak24}                                                                                                                         \\ 
\hline
DnCNN-B\cite{zhang2017beyond}                       & 0.764                        & 0.713                        & 0.675                        & 0.631                  & ——                             \\ 
\hline
ADNet\cite{tian2020attention}                         & \textcolor{blue}{0.768}                        & \textcolor{blue}{0.717}                        & \textcolor{blue}{0.678}                        & \textcolor{blue}{0.635}                  & \textcolor{blue}{0.585}                          \\ 
\hline
FFDNet\cite{zhang2018ffdnet}                        & 0.766                        & 0.709                        & 0.668                        & 0.621                  & 0.570                          \\
\hline
MWDCNN (Ours)                            & \textcolor{red}{0.773}       & \textcolor{red}{0.722}       & \textcolor{red}{0.686}       & \textcolor{red}{0.643} & \textcolor{red}{0.596}         \\ 
\hline
\end{tabular}
\end{table}
\label{tab:booktabs}

%%%%%%%%%%%%%%%%%%%%%%%%%%%%%%%%%%%%%%%%%%%%%%%%%%%%%%%%%%%%%%%%%%%%%%%%%%%%%%%%%%%%%%%%%%%%%%%%%%% %%%%%%%%%%%%%%%%%%%%%%%%%%%%%%%%%%%%%%%%%%%%Table12%%%%%%%%%%%%%%%%%%%%%%%%%%%%%%%%%%%%%%%%%%%%%%%%% %%%%%%%%%%%%%%%%%%%%%%%%%%%%%%%%%%%%%%%%%%%%%%%%%%%%%%%%%%%%%%%%%%%%%%%%%%%%%%%%%%%%%%%%%%%%%%%%%%%

% \usepackage{color}

\begin{table}
\caption{SSIM results of different methods on Kodak24 with noise levels of 15, 25, 35, 50 and 75.}
\centering
\begin{tabular}{|l|l|l|l|l|l|} 
\hline
 Noise Level   & $\sigma=15$                & $\sigma=25$                & $\sigma=35$                & $\sigma=50$                & $\sigma=75$                 \\ 
\hline
 DnCNN-B\cite{zhang2017beyond}       &0.9205  & 0.8774 &0.8394 &0.7896 & ——       \\ 
\hline
 ADNet\cite{tian2020attention}         &\textcolor{blue}{0.9239}  &\textcolor{blue}{0.8820}  &\textcolor{blue}{0.8445} &\textcolor{blue}{0.7984}  &\textcolor{blue}{0.7387}  \\ 
\hline
 FFDNet \cite{zhang2018ffdnet}        & 0.9231  &  0.8792 &0.8409  &0.7930 & 0.7328  \\ 
\hline
 MWDCNN (Ours)  &\textcolor{red}{0.9269}   &\textcolor{red}{0.8862}   &\textcolor{red}{0.8515}   &\textcolor{red}{0.8062}   &\textcolor{red}{0.7491}    \\
\hline
\end{tabular}
\end{table}

%%%%%%%%%%%%%%%%%%%%%%%%%%%%%%%%%%%%%%%%%%%%%%%%%%%%%%%%%%%%%%%%%%%%%%%%%%%%%%%%%%%%%%%%%%%%%%%%%%% %%%%%%%%%%%%%%%%%%%%%%%%%%%%%%%%%%%%%%%%%%%%Table13%%%%%%%%%%%%%%%%%%%%%%%%%%%%%%%%%%%%%%%%%%%%%%%%% %%%%%%%%%%%%%%%%%%%%%%%%%%%%%%%%%%%%%%%%%%%%%%%%%%%%%%%%%%%%%%%%%%%%%%%%%%%%%%%%%%%%%%%%%%%%%%%%%%%

% \usepackage{color}

\begin{table}
\caption{LPIPS results of different methods on Kodak24 with noise levels of 15, 25, 35, 50 and 75.}
\centering
\begin{tabular}{|l|l|l|l|l|l|} 
\hline
 Noise Level  & $\sigma=15$                & $\sigma=25$                & $\sigma=35$                & $\sigma=50$                & $\sigma=75$                 \\ 
\hline
DnCNN-B \cite{zhang2017beyond} &  0.1719 & 0.2323 & 0.2780 & 0.3344 & ——      \\ 
\hline
 ADNet \cite{tian2020attention}    & \textcolor{blue}{0.1663} &\textcolor{blue}{0.2256}  &\textcolor{blue}{0.2690} &\textcolor{blue}{0.3228}  &\textcolor{blue}{0.3837}  \\ 
\hline
FFDNet \cite{zhang2018ffdnet}   &  0.1721 & 0.2376  & 0.2866 &0.3408 &0.4055  \\ 
\hline
 MWDCNN (Ours)  &\textcolor{red}{0.1567}   &\textcolor{red}{0.2154}   &\textcolor{red}{0.2579}   &\textcolor{red}{0.3075}   &\textcolor{red}{0.3639}    \\
\hline
\end{tabular}
\end{table}

%%%%%%%%%%%%%%%%%%%%%%%%%%%%%%%%%%%%%%%%%%%%%%%%%%%%%%%%%%%%%%%%%%%%%%%%%%%%%%%%%%%%%%%%%%%%%%%%%%% %%%%%%%%%%%%%%%%%%%%%%%%%%%%%%%%%%%%%%%%%%%%Table14%%%%%%%%%%%%%%%%%%%%%%%%%%%%%%%%%%%%%%%%%%%%%%%%% %%%%%%%%%%%%%%%%%%%%%%%%%%%%%%%%%%%%%%%%%%%%%%%%%%%%%%%%%%%%%%%%%%%%%%%%%%%%%%%%%%%%%%%%%%%%%%%%%%%

% \usepackage{color}

\begin{table}
\caption{ Inception-Score of different methods on Kodak24 with noise levels of 15 and 50.}
\centering
\begin{tabular}{|l|l|l|} 
\hline
Noise Level  & $\sigma=15$              & $\sigma=50$               \\ 
\hline
 DnCNN-B \cite{zhang2017beyond}  &\textcolor{blue}{9.68} &8.97  \\ 
\hline
 ADNet \cite{tian2020attention}    &9.41 &\textcolor{blue}{9.31}  \\ 
\hline
FFDNet \cite{zhang2018ffdnet}   &9.46 & 9.21  \\ 
\hline
 MWDCNN (Ours) &\textcolor{red}{9.83}   &\textcolor{red}{9.52}    \\
\hline
\end{tabular}
\end{table}

%%%%%%%%%%%%%%%%%%%%%%%%%%%%%%%%%%Fig4%%%%%%%%%%%%%%%%%%%%%%%%%%%%%%%%%%%%%%%%%%%%%
\begin{figure*}[!t]
\centering
\subfloat{\includegraphics[width=5.0in]{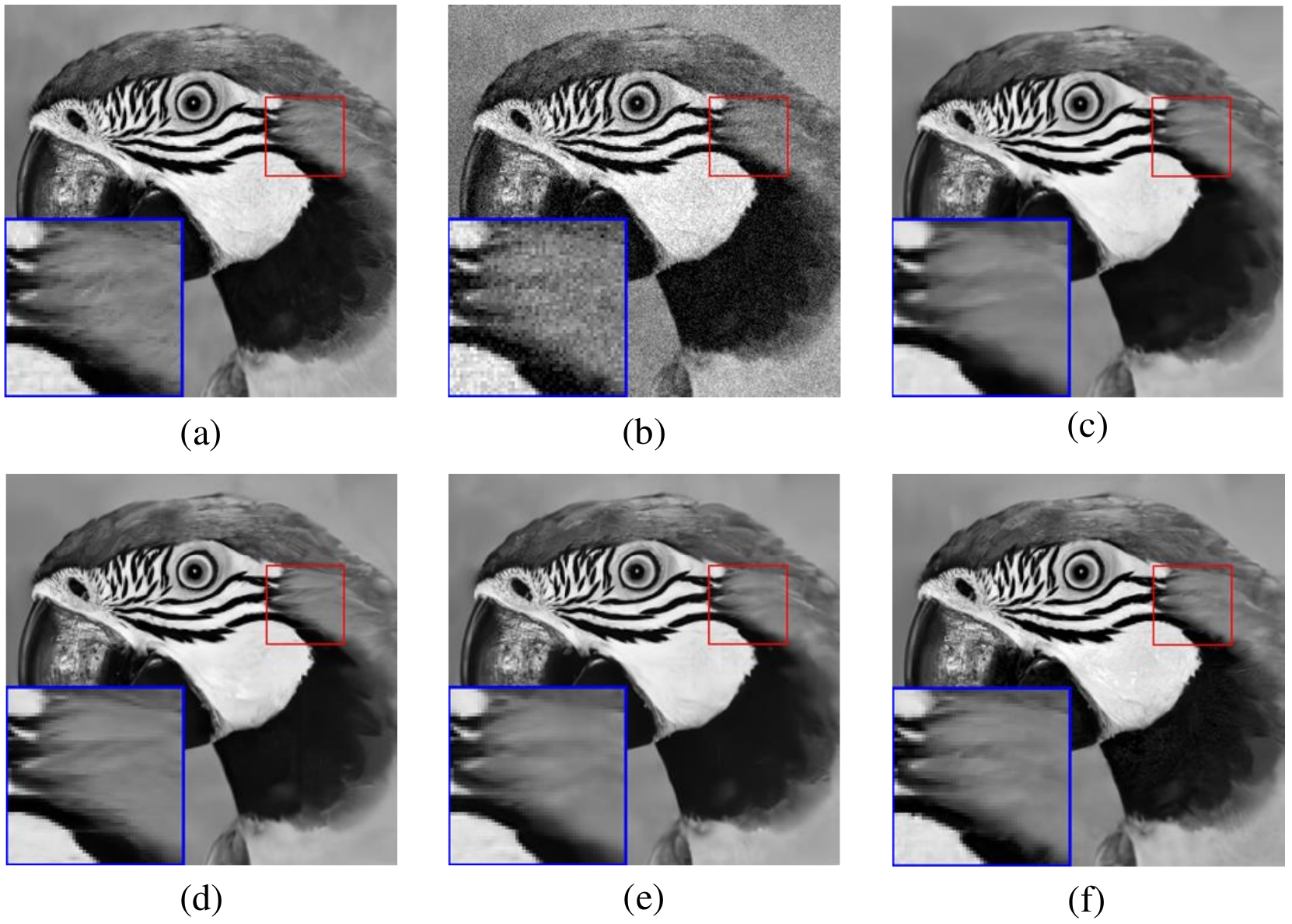}
\label{fig_second_case}}
\caption{Denoising results of a gray noisy image from Set12 with ${\sigma}$ = 15. (a) Clean image, (b) Noisy image, (c) FFDNet/31.81dB, (d) ADNet/31.96dB, (e) DnCNN/31.83dB and (f) MWDCNN (Ours)/31.97dB.}
\label{fig:7}
\end{figure*}

%%%%%%%%%%%%%%%%%%%%%%%%%%%%%%%%%%Fig5%%%%%%%%%%%%%%%%%%%%%%%%%%%%%%%%%%%%%%%%%%%%%
\begin{figure*}[!t]
\centering
\subfloat{\includegraphics[width=5.0in]{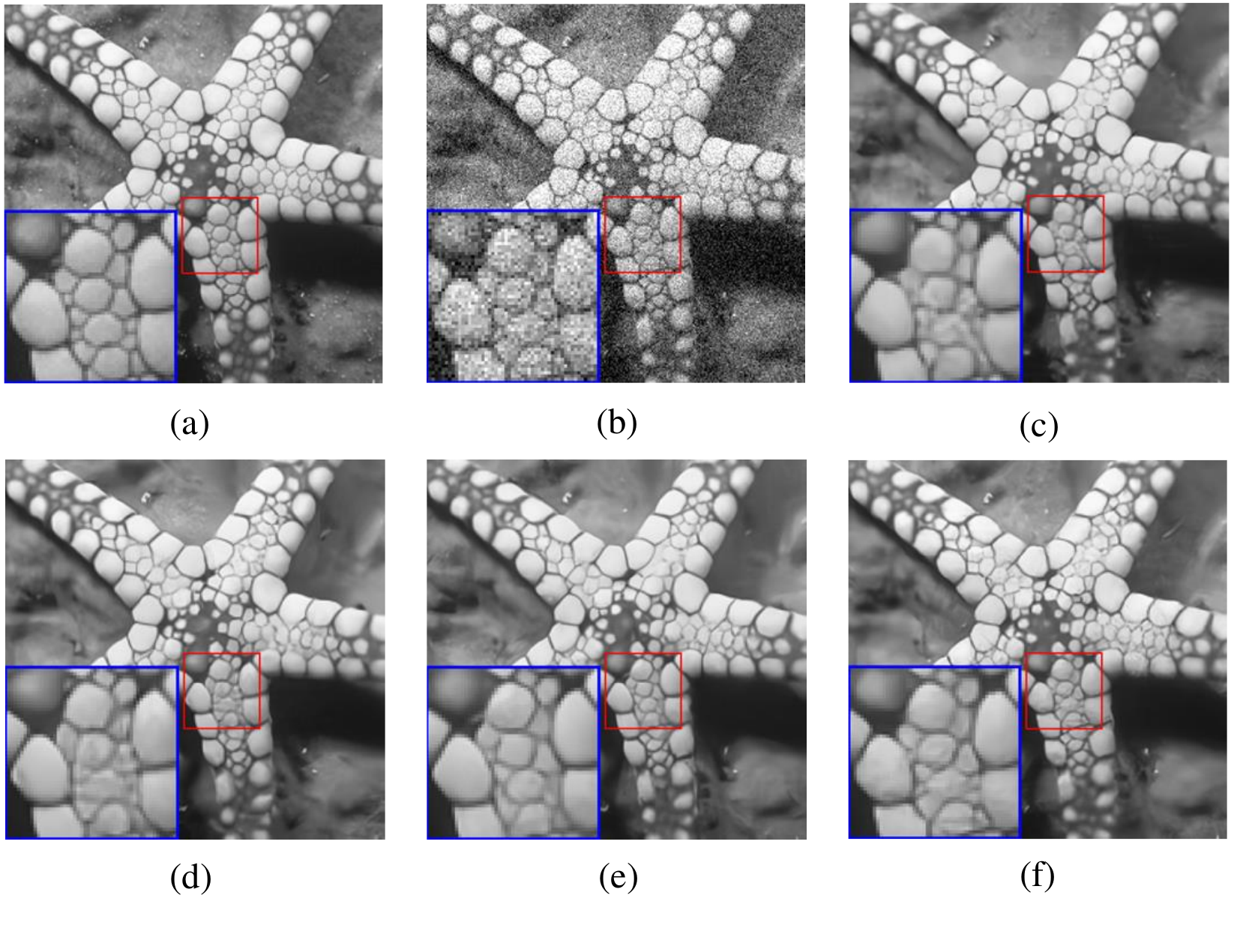}
\label{fig_second_case}}
\caption{Denoising results of a gray noisy image from Set12 with ${\sigma}$ = 25. (a) Clean image, (b) Noisy image, (c) FFDNet/29.32dB, (d) ADNet/29.41dB,  (e) DnCNN/29.41dB and  (f) MWDCNN (Ours)/29.66dB.}
\label{fig:7}
\end{figure*}

%%%%%%%%%%%%%%%%%%%%%%%%%%%%%%%%%%Fig6%%%%%%%%%%%%%%%%%%%%%%%%%%%%%%%%%%%%%%%%%%%%%
\begin{figure*}[!t]
\centering
\subfloat{\includegraphics[width=5.0in]{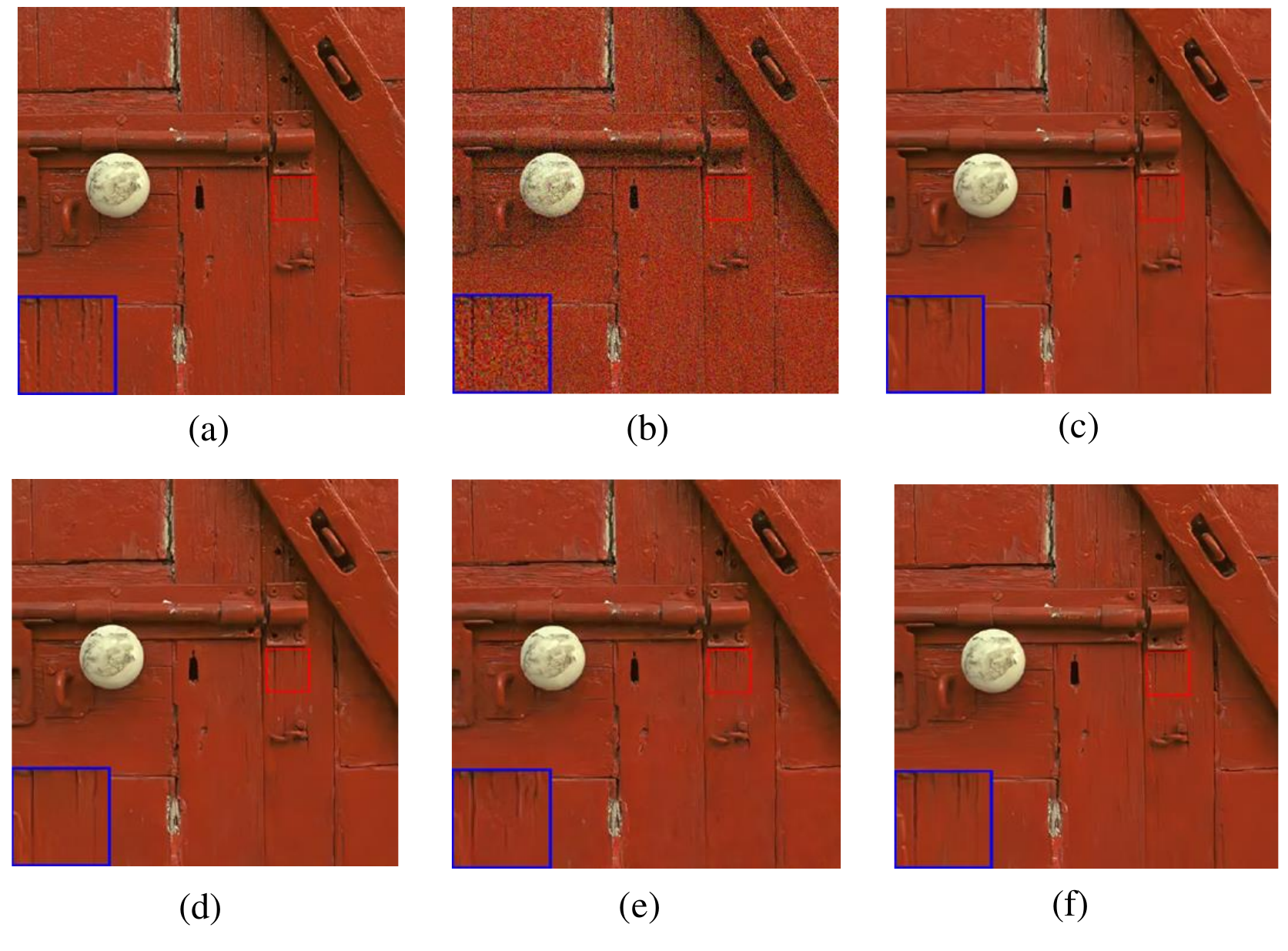}
\label{fig_second_case}}
\caption{Denoising results of a color noisy image from Kodak 24 with ${\sigma}$ = 25. (a) Clean image, (b) Noisy image, (c) FFDNet/32.59dB, (d)ADNet/32.30dB, (e) DnCNN/32.53dB and (f) MWDCNN (Ours)/32.82dB.}
\label{fig:7}
\end{figure*}

%%%%%%%%%%%%%%%%%%%%%%%%%%%%%%%%%%Fig7%%%%%%%%%%%%%%%%%%%%%%%%%%%%%%%%%%%%%%%%%%%%%
\begin{figure*}[!t]
\centering
\subfloat{\includegraphics[width=5.0in]{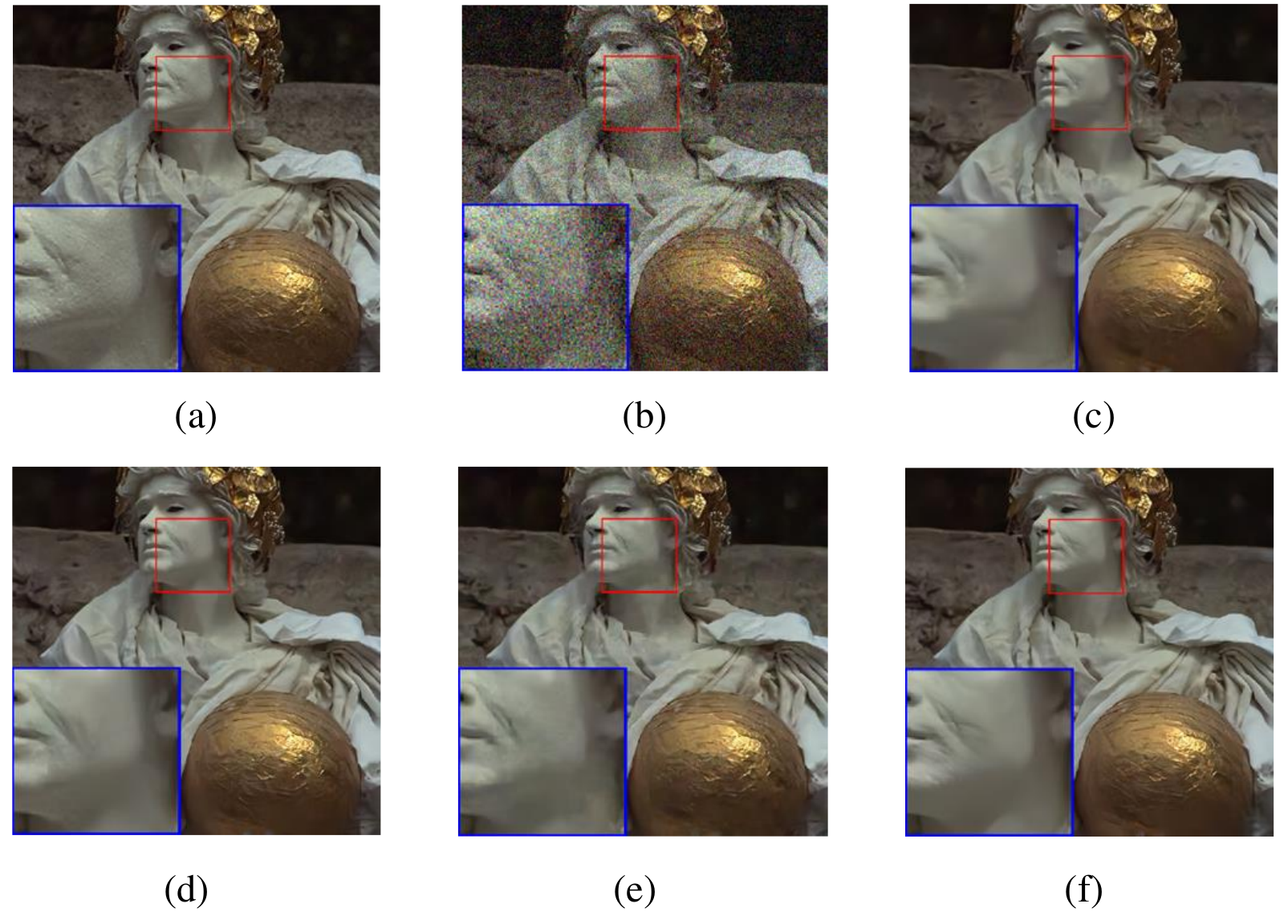}
\label{fig_second_case}}
\caption{Denoising results of a color noisy image from Kodak 24 with ${\sigma}$ = 35. (a) Clean image, (b) Noisy image, (c) FFDNet/31.05dB, (d) ADNet/30.24dB, (e) DnCNN/30.93dB and (f) MWDCNN (Ours)/31.20dB.}
\label{fig:7}
\end{figure*}

\subsection{Comparisons with state-of-the-art denoising methods}

In this section, we quantitatively and qualitatively analyze denoising performance of the proposed MWDCNN. Specifically, quantitative analysis is evaluated by comparing PSNR, feature similarity index measure (FSIM)\cite{zhang2011fsim},  structural similarity  index measure (SSIM)\cite{wang2004image}, learned perceptual image patch similarity (LPIPS)  as well as DeepFeatures\cite{zhang2018unreasonable} and Inception-Score\cite{salimans2016improved} of popular denoising methods containing  DnCNN\cite{zhang2017beyond},  block-matching and 3-D filtering (BM3D) \cite{dabov2007image}, FFDNet\cite{zhang2018ffdnet}, ADNet\cite{tian2020attention}, dual denoising network (DudeNet)\cite{tian2021designing}, weighted nuclear norm minimization method (WNNM)\cite{gu2014weighted}, trainable nonlinear reaction diffusion (TNRD)\cite{chen2016trainable}, complex-valued denoising network (CDNet)\cite{quan2021image}, grey theory applied in non-local means (GNLM)\cite{li2016novel}, image gradient network (GradNet)\cite{liu2020gradnet} and targeted image denoising (TID)\cite{luo2015adaptive} on some public datasets, i.e., BSD68\cite{li2013benchmark}, Set12\cite{li2013benchmark}, CBSD68\cite{li2013benchmark}, Kodaka24\cite{franzen1999kodak} for different noise levels and CC\cite{nam2016holistic} for real noisy image denoising. For gray Gaussian noisy image denoising, we use BSD68 and Set12 to test denoising performance of the proposed MWDCNN as shown in Tables 5 and 6. In Table 5, we can see that MWDCNN is superior to DnCNN for three different noise levels (i.e., 15, 25 and 50) on BSD68. MWDCNN-B as well as MWDCNN for blind denoising is conducted via varying noise levels from 0 to 55. Besides, the proposed MWDCNN has obtained excellent denoising results for each noisy images on the Set12 in PSNR for three noise levels (i.e., 15, 25 and 50) as given in Table 6. 
     
For color Gaussian noisy image denoising, we use CBSD68 and Kodak24 to test denoising performance of different methods for noise levels of 15, 25, 35, 50 and 75 as shown in Tables 7 and 8, respectively. For instance, our MWDCNN as the best result exceeds ADNet as the second result in PSNR on CBSD68 for mentioned five noise levels in Table 7. Also, our MWDCNN is more effective than these of other comparative denoising methods on Kodak24 for five noise levels above in Table 8. These illustrations our MWDCNN is a good tool for color Gaussian noisy image denoising. 

For real noisy image denoising, we use CC to test denoising performance of different methods for real digital devices in the real world. As shown in Table 9, our MWDCNN has obtained higher average PSNR than that of other denoising methods, which shows our MWDCNN is effective in real noisy image denoising.

To further test denoising performance, we use FSIM\cite{zhang2011fsim} on BSD68 for noise levels of 15, 25 and 50, CBSD68 and Kodak24 for noise levels of 15, 25, 35, 50 and 75, and SSIM\cite{wang2004image}  and LPIPSM\cite{zhang2018unreasonable} on Kodak24 for noise levels of 15, 25, 35, 50 and 75 to conduct experiments in terms of low-level perceptual aspects. As shown Tables 10 and 11, we can see that our MWDCNN obtained the highest FSIM for all noise levels to verify its excellent denoising performance. As shown in Tables 12 and 13, we can see that our MWDCNN is superior to other popular denoising methods, i.e., DnCNN-B, ADNet and FFDNet in terms of SSIM and LPIPS. In terms of high-level metrics, we choose Inception-Score\cite{salimans2016improved} 
to conduct experiments as shown in Table 14, which shows our methods is more effective than DnCNN-B, ADNet and FFDNet in low- and high-noise levels. The best denoising results are marked by red and blue lines from Table 5 to Table 14..  

Due to difficult improvements of denoising methods, improvement of 0.01dB from denoisers are very reasonable \cite{tian2020deep,zhang2017beyond}. Thus, although some MWDCNN have only  a little improvements of PSNR than that of the second image denoising methods for certain noise level for gray and color Gaussian noisy image denoising, they are very competitive for image denoising. Also, the proposed MWDCNN is very effective for real noisy image denoising. Besides,
our MWDCNN is beneficial to low-level metrics, i.e., FSIM, SSIM and LPIPS and high-level metric, i.e., Inception-Score. These illustrate the proposed MWDCNN is effective and robust in image denoising in terms of quantitative analysis.
   
In terms of qualitatively analysis, we choose popular denoising methods, i.e.,  FFDNet,  ADNet, DnCNN and our MWDCNN to obtain predicted images to compare visual denoising performance. The predicted images of different methods can be chosen one area to amplify this area as observation area, the mentioned observation area is clearer, its corresponding denoising method is more effective for image denoising. For gray noisy image denoising, we choose two images from the Set12 to conduct visual experiments for noise levels of 15 and 25, respectively. As shown in Fig. 4 and Fig. 5, we can see that our MWDCNN is clearer than these of other denoising methods in the observation areas. That shows it is more effective in visual perspective for gray image denoising. For color noisy image denoising, we choose two images from the Kodak24 to conduct visual experiments for noise levels of 25 and 35, respectively. As shown in Fig. 6 and Fig. 7, we can see that our MWDCNN is clearer than other denoising methods in the observation areas. That illustrates it is more effective in perspective for color image denoising. According to mentioned descriptions, we can see that the proposed MWDCNN is more superior than other popular denoising methods in qualitatively analysis. In a summary, our MWDCNN is a good choice for image denoising, according to quantitative and qualitative analysis. 

According to analysis of Sections 4.3 and 4.4, we can see that due to use of a dynamic convolution, the proposed MWDCNN is strong adaptability and computational costs  for different scenes. Due to combination of a single processing technique and discriminative learning technique as well as fusing wavelet transform technique and a CNN, the proposed MWDCNN has better denoising performance. Due to use of stacked architectures, our MWDCNN is robust for image denoising. Thus, the proposed MWDCNN is effective to complex scenes for portable digital devices, i.e., smartphones and cameras. 
\section{Conclusion}
In this paper, we propose a multi-stage image denoising CNN with the wavelet transform as well as MWDCNN in image denoising. The first stage of MWDCNN uses linearly combination of several convolutional kernels to dynamically adjust parameters rather than same parameters of convolutions, according to different noisy images, which can make a tradeoff between denoising performance and computational cost. It is very useful for noisy images with unknown noise. The second stage of MWDCNN fuses frequency features and structure features via two combinations of wavelet transformation and residual dense block to suppress noise, which enhances robustness of obtained denoiser for complex scenes. The third stage of MWDCNN uses improved residual dense architectures to remove redundant features to improve performance of image denoising. The proposed method is very beneficial to damaged images with unknown noise. However, it depends on a supervised manner to train a denoising model. Also, collected noisy images by cameras are difficult to obtain clean reference  images. Thus, we will conduct clean reference images to train a blind denoiser in the future, according to properties of noisy images. 

\section*{Acknowledgments}
This work was supported in part by National Natural Science Foundation of China under Grant 62201468, in part by the China Postdoctoral Science Foundation Grant 2022TQ0259, in part by the Jiangsu Provincial Double–Innovation Doctor Program Grant JSSCBC20220942.
%\section*{References}
%% References


% \bibliographystyle{elsarticle-num}
\bibliographystyle{elsarticle-harv}
\bibliography{references}
%\bibliography{sample}

%-----------------------------------------------------------------------------------------------------
\end{spacing}
\end{document}